\documentclass
[preprint,aps,onecolumn,titlepage,superscriptaddress,letterpaper,10pt]{revtex4}%
\usepackage{amsmath}
\usepackage{amssymb}
\usepackage{graphicx}
\usepackage{amsfonts}%
\begin{document}
\title[RGFM]{Renormalization Group Functional Equations}
\author{Thomas L. Curtright}
\affiliation{CERN, CH-1211 Geneva 23, Switzerland}
\affiliation{Department of Physics, University of Miami, Coral Gables, FL 33124-8046, USA}
\author{Cosmas K. Zachos}
\affiliation{High Energy Physics Division, Argonne National Laboratory, Argonne, IL
60439-4815, USA\medskip\medskip}

\begin{abstract}
Functional conjugation methods are used to analyze the global structure of
various renormalization group trajectories, and to gain insight into the
interplay between continuous and discrete rescaling. \ With minimal
assumptions, the methods produce continuous flows from step-scaling $\sigma$
functions, and lead to exact functional relations for the local flow $\beta$
functions, whose solutions may have novel, exotic features, including multiple
branches. \ As a result, fixed points of $\sigma$ are sometimes \emph{not}
true fixed points under continuous changes in scale, and zeroes of $\beta$ do
\emph{not} necessarily signal fixed points of the flow, but instead may only
indicate turning points of the trajectories.

\end{abstract}
\preprint{ANL-HEP-PR-10-52, CERN-PH-TH-2010-245}
\volumeyear{year}
\volumenumber{number}
\issuenumber{number}
\eid{identifier}
\startpage{1}
\endpage{ }
\maketitle

\section{Introduction}

The renormalization group (RG) of Gell-Mann and Low \cite{GML}, and of
Stueckelberg and Petermann \cite{SP}, has an elegant mathematical expression
in terms of the functional conjugation (FC) methods of Ernst Schr\"{o}der
\cite{S}. \ This expression provides a powerful tool to describe the behavior
of physical systems under either infinitesimal or finite, perhaps large,
changes in scale. \ While this fact is often overlooked, and not usually
invoked in the solution of various problems posed in the RG framework, it is
readily apparent upon reading \cite{GML} (see especially Appendix B; also see
\cite{Lee}) and surveying the literature on functional equations \cite{Ku}.
\ Moreover, it may be profitable to bear in mind the logical connections
between these two subjects when considering the step-scaling approach in
lattice gauge theory \cite{C,L}, where the power and utility of the methods
are manifest.

In previous work \cite{CZ1, CZ2, CV}\ we have discussed how dynamical systems,
defined on a discrete lattice of time points, may be smoothly interpolated in
time through the use of solutions to Schr\"{o}der's celebrated functional
equation. \ Here we discuss the same methods in the context of the
renormalization group. \ We examine in detail the connections between
differential (local) rescaling and finite (global) changes in scale. \ We
interpolate various step-scaling functions to obtain trajectories under
continuous change of scale, with emphasis on the consistency imposed by the
analytic properties of couplings in the presence of UV and IR\ fixed points.
\ From this point of view it is possible to obtain novel features for RG
behavior. \ In particular, multi-valued Callan--Symanzik $\beta$ functions
\cite{Ca,GML,SP,Sy} are commonly encountered in the local RG flow equations,
even when interpolating elementary, polynomial step-scaling functions, with
interesting consequences involving fixed points, cycles, and even chaotic
evolution under changes in scale.

In section II we describe functional conjugation methods relevant to RG
analysis and apply them to the study of selected trajectories. \ In section
III we consider a physical illustration of fixed point behavior drawn from
numerical studies of lattice gauge theory \cite{A}. \ In section IV we
illustrate elementary limit cycle behavior in a model obtained by an extension
of the standard BCS Hamiltonian \cite{LeC}. \ In section V we briefly explain
how further novel, exotic features can arise from basic step-scaling behavior,
in general. \ Finally, in section VI we exhibit such features, including
multi-valued $\beta$ functions, limit cycles, and chaotic trajectories, using
toy models based on the logistic map. \ Two appendices provide some
connections to our earlier work on dynamical systems, and a few algebraic
details for the lattice example.

\section{Methodology}

\subsection{The renormalization group: \ Step by step}

Let us suppose the change in the coupling\ $u$ is given for a discrete change
in length scale by%
\begin{equation}
u\mapsto\sigma\left(  u\right)  \ , \label{StepScalingMap}%
\end{equation}
where $\sigma\left(  u\right)  $ is the \textquotedblleft
step-scaling\textquotedblright\ function \cite{L,C}. \ Typically,
$u=g^{2}/4\pi$, where $g$ is the gauge coupling, although it may be convenient
to incorporate other numerical factors into $u$, or even to take other
functions of $g^{2}$, depending on the problem at hand. \ The standard
interpretation is to regard $\sigma\left(  u\right)  $ as a discrete $\Delta
t$ sampling of a renormalization trajectory, $u\left(  t\right)  $, whose
continuous\ evolution under changes in the \emph{log} of the length scale, $t
$, has proceeded from an initial $u\equiv\left.  u\left(  t\right)
\right\vert _{t=0}$. \ This notation for the initial $u$ (rather than $u_{0}$,
say) is not only more convenient to express the step-scaling function, e.g. as
the mapping (\ref{StepScalingMap}), but also to write many relations that hold
both for the initial $u$ as well as more generally for all $u\left(  t\right)
$. \ We will point out several such relations in the following.

The trajectory is assumed to describe an abelian $t$-flow with group
composition given by simple addition of $t$ arguments. \ So, for example,
$\sigma\left(  u\right)  =\left.  u\left(  t\right)  \right\vert _{t=1}$,
$\sigma\left(  \sigma\left(  u\right)  \right)  =\left.  u\left(  t\right)
\right\vert _{t=2}$, $\sigma^{-1}\left(  u\right)  =\left.  u\left(  t\right)
\right\vert _{t=-1}$, etc. \ The local flow equation in terms of $t$ has both
familiar and more recondite forms,
\begin{equation}
\frac{du}{dt}=\beta\left(  u\right)  \equiv\left(  \ln\lambda\right)
\Psi\left(  u\right)  /\Psi^{\prime}\left(  u\right)  , \label{beta}%
\end{equation}
where $\beta$ is the so-called Callan-Symanzik function \cite{Ca,Sy} (which
appeared earlier in \cite{SP}, Eq(4.25), under the alias $h_{i\rho}$) and
$\Psi$ is the so-called Schr\"{o}der function \cite{S} (both of which appeared
in \cite{GML}, under the aliases $\psi$ and $G$, respectively), and where
$1/\ln\lambda$ (usually taken to be $\pm1$) sets the scale of $t$. \ It is
also implicitly understood that the system is underlain by a $t$-translation
covariance so that (\ref{beta})\ holds not just for $u\equiv\left.  u\left(
t\right)  \right\vert _{t=0}$\ but also for $u\left(  t\right)  $, provided of
course that the RHS is also modified by $\beta\left(  u\right)  \rightarrow
\beta\left(  u\left(  t\right)  \right)  $ and $\Psi\left(  u\right)
\rightarrow\Psi\left(  u\left(  t\right)  \right)  $.

While the Schr\"{o}der function is less well-known in renormalization theory,
it is immediately expressed in terms of $\beta$ from the definition in
(\ref{beta}), rewritten as
\begin{equation}
d\ln\Psi\left(  u\right)  /du=\left(  \ln\lambda\right)  /\beta\left(
u\right)  \ . \label{ln(Psi)}%
\end{equation}
Thus a definite integral gives the total change in $\Psi$ brought about by a
finite change in the coupling,%
\begin{equation}
\Psi\left(  u_{2}\right)  =\lambda^{\int_{u_{1}}^{u_{2}}\frac{d\mathfrak{u}%
}{\beta\left(  \mathfrak{u}\right)  }}~\Psi\left(  u_{1}\right)  \ .
\label{PsiFiniteStep}%
\end{equation}
On the other hand, the exponent here is just $t_{2}-t_{1}$, the total change
in $t$ as $u_{1}\rightarrow u_{2}$, as follows from the first equality in
(\ref{beta}). \ Therefore another way to express (\ref{PsiFiniteStep}) is in
terms of the evolution of the Schr\"{o}der function under the flow of the
coupling, $u\rightarrow u\left(  t\right)  $, \
\begin{equation}
\Psi\left(  u\left(  t\right)  \right)  =\lambda^{t}~\Psi\left(  u\right)  \ .
\label{Psi(u(t))}%
\end{equation}
This last relation reveals the fundamental role played by $\Psi$, and its
inverse function $\Psi^{-1}$, in the construction of trajectories for
arbitrary changes in $t$. \ It follows from (\ref{Psi(u(t))}) that such
\emph{global} \emph{flow} is given by \cite{GML}%
\begin{equation}
u\left(  t\right)  =\Psi^{-1}\left(  \lambda^{t}\Psi\left(  u\right)  \right)
\ , \label{global}%
\end{equation}
where $1/\ln\lambda$ sets the scale of $t$. \ The RHS of (\ref{global}) is
immediately recognized as just a change of variable, effected through a
functional conjugation \cite{S}. \ For us, in fact, the expression
\textquotedblleft Schr\"{o}der functional method\textquotedblright\ is just a
metonymy for \emph{functional conjugation}. \ 

Indeed, the expression (\ref{global}) is perhaps the most succinct way to
appreciate that renormalization relates self-similar structures at different
scales, inasmuch as the RHS is just a functional similarity transformation:
\ $\Psi^{-1}\circ\lambda^{t}\circ\Psi$. \ 

Moreover, (\ref{global}) shows that fixed points or limit cycles can arise in
a model for real $\lambda$ if and only if $\Psi^{-1}$ either becomes constant
or else exhibits periodic behavior, respectively.

The structure of (\ref{global}) also makes the abelian $t$-flow of the
renormalization group manifest, and it gives a formula for the step-scaling
function, or any of its functional compositions, in terms of $\Psi$. \ For
example, for $\sigma\left(  u\right)  \equiv\left.  u\left(  t\right)
\right\vert _{t=1}$, we have from (\ref{Psi(u(t))})
\begin{equation}
\lambda\Psi\left(  u\right)  =\Psi\left(  \sigma\left(  u\right)  \right)  \ ,
\label{Schroeder}%
\end{equation}
a form known as \textquotedblleft Schr\"{o}der's functional equation with
eigenvalue $\lambda$.\textquotedblright\ \ Presented in this form, for a given
$\sigma\left(  u\right)  $, the problem is often to determine all allowed
$\lambda$ and to find all solutions of the functional equation
\cite{Ambiguity}. \ 

Alternatively, we may write (\ref{Schroeder}) as
\begin{equation}
\sigma\left(  u\right)  =\Psi^{-1}\left(  \lambda\Psi\left(  u\right)
\right)  \ .
\end{equation}
In this form, the equation determines the step-scaling function in terms of
$\Psi$. \ In fact, it is useful to think of $u\left(  t\right)  $ in
(\ref{global}) as $\sigma_{t}\left(  u\right)  $, that is, as a $t$-th
\emph{continuous} functional composition of $\sigma$. \ For example,
$\sigma_{2}\left(  u\right)  =\sigma\left(  \sigma\left(  u\right)  \right)  $
as before, but now generalized to $\sigma\left(  u\right)  =\sigma
_{1/2}\left(  \sigma_{1/2}\left(  u\right)  \right)  $, etc. \ More generally,
$\sigma_{t_{1}+t_{2}}\left(  u\right)  =\sigma_{t_{1}}\left(  \sigma_{t_{2}%
}\left(  u\right)  \right)  $ --- just the expected RG abelian composition rule.

At this point it is natural to ask, what is a simple physical model whereby
$\Psi\left(  t\right)  =\lambda^{t}\Psi_{0}$? \ Well, $d\ln\Psi\left(
t\right)  /dt=\ln\lambda$, so clearly $\ln\Psi$ is the variable of choice.
\ Then the question becomes, for what model is the change in the coupling with
scale a constant? \ An obvious answer is, \emph{the one-loop approximation}
for evolution of an inverted coupling, $1/\mathfrak{g}^{2}$. \ That is to say,
if
\begin{equation}
\frac{d}{dt}\mathfrak{g}\left(  t\right)  =\beta_{\mathfrak{g}\text{ 1-loop}%
}=\frac{1}{2}\mathfrak{c}\mathfrak{g}^{3}\left(  t\right)  \ ,
\end{equation}
then
\begin{equation}
\frac{d}{dt}\left(  \frac{1}{\mathfrak{g}^{2}\left(  t\right)  }\right)
=-\mathfrak{c}\ .
\end{equation}
So, the physical interpretation of the Schr\"{o}der function is clear: \ The
log of $\Psi$ is just the change of variable needed to convert the
renormalization group flow for $u$ into a one-loop flow for a re-defined
coupling constant $1/\mathfrak{g}^{2}$. \ Thus,%
\begin{equation}
\frac{d}{dt}\ln\Psi\left(  u\left(  t\right)  \right)  =\ln\lambda
\ \ \ \Longleftrightarrow\ \ \ \frac{d}{dt}\left(  \frac{1}{\mathfrak{g}%
^{2}\left(  t\right)  }\right)  =-\mathfrak{c}\ .
\end{equation}

The role of $\ln\Psi$ is put into deeper perspective through the following
formal sequence of steps that evoke the method of characteristics for the RG.
\ Making explicit the dependence of the trajectory on the initial $u=\left.
u\left(  t\right)  \right\vert _{t=0}$ as well as on $t$, and making use of
\begin{equation}
\frac{\partial}{\partial t}=\beta\left(  u\right)  \frac{\partial}{\partial
u}=\left(  \ln\lambda\right)  \frac{\partial}{\partial\ln\Psi\left(  u\right)
} \label{characteristics}%
\end{equation}
along the trajectory, we have%
\begin{equation}
u\left(  t,u\right)  =\left.  e^{t\frac{\partial}{\partial\tau}}~u\left(
\tau,u\right)  \right\vert _{\tau=0}=e^{t\beta\left(  u\right)  \frac
{\partial}{\partial u}}~u=e^{t\left(  \ln\lambda\right)  \frac{\partial
}{\partial\ln\Psi\left(  u\right)  }}~u\ .
\end{equation}
But now, $u=\Psi^{-1}\left(  \Psi\left(  u\right)  \right)  =\Psi^{-1}\left(
\exp\left(  \ln\Psi\left(  u\right)  \right)  \right)  $, so the last
expression reduces to a mere translation of the variable $\ln\left(
\Psi\left(  u\right)  \right)  $,
\begin{equation}
e^{t\left(  \ln\lambda\right)  \frac{\partial}{\partial\ln\Psi\left(
u\right)  }}~\Psi^{-1}\left(  \exp\left(  \ln\Psi\left(  u\right)  \right)
\right)  =\Psi^{-1}\left(  \exp\left(  t\ln\lambda+\ln\Psi\left(  u\right)
\right)  \right)  =\Psi^{-1}\left(  \lambda^{t}\Psi\left(  u\right)  \right)
\ .
\end{equation}
Thus (\ref{global}) is recovered. \ These formal steps can be made precise by
examination of $T$, the radius of convergence of the $t$ series, in particular
by a determination of the dependence of $T$ on the initial $u$. \ How this
goes will be illustrated in the examples to follow.

There is some additional, subtle mathematical structure to take into account
here, especially if we have in hand a series expansion for $\sigma$:
\begin{equation}
\sigma\left(  u\right)  =\alpha u+O\left(  u^{2}\right)  \ .
\label{SigmaSeries}%
\end{equation}
For example, if $\frac{d}{dt}u\left(  t\right)  =cu^{2}\left(  t\right)  $, as
is true for lowest order perturbation theory with $u\propto g^{2}$, then the
exact solution for the trajectory is
\begin{equation}
u\left(  t\right)  =\frac{u}{1-cut}\ ,
\end{equation}
where again on the RHS $u=\left.  u\left(  t\right)  \right\vert _{t=0}$. \ In
this case we have $\alpha=1$ in (\ref{SigmaSeries}). \ 

But this leads to $\lambda=1$, the well-known singular situation for the
eigenvalue in Schr\"{o}der's functional equation. \ By singular we mean that
an attempt to solve (\ref{Schroeder}) by Taylor series expansion about $u=0$
will fail, in general, when $\lambda=1$. \ Indeed, it is immediately verified
that, for $\alpha=1$ in (\ref{SigmaSeries}), a nontrivial solution of
(\ref{Schroeder}) can \emph{not} be found with $\Psi\left(  u\right)  $ given
by a series about $u=0$. \ 

This is easily circumvented, however. \ Instead of Taylor series about $u=0$,
all the relevant solutions have an essential singularity at $u=0$, and in fact
have Taylor series about $u=\infty$. \ Explicitly, with $\lambda\equiv
\exp\left(  \kappa c\right)  $ in (\ref{ln(Psi)}), we see that $\Psi_{\kappa
}\left(  u\right)  =\exp\left(  -\frac{\kappa}{u}\right)  $ is a family of
Schr\"{o}der functions for one-loop evolution, with arbitrary $\kappa$:
\ $\Psi_{\kappa}\left(  u\left(  t\right)  \right)  =e^{\kappa ct}%
~\Psi_{\kappa}\left(  u\right)  $. \ So, for $t=1$ and $\sigma\left(
u\right)  \equiv\left.  u\left(  t\right)  \right\vert _{t=1}$, the eigenvalue
for each $\Psi_{\kappa}$ solution is indeed $\lambda$, and not the naive value
$1$\ (excluding the trivial and uninteresting case where $\Psi_{\kappa=0}=1$).
\ Defining a discrete step for another value of $t$ simply rescales $\kappa$. \ 

We will say more about the general structure of the functional approach to the
RG, and the novel features that it has the power to reveal, in sections IV and
V of the paper. \ But first, we consider:

\subsection{The $\beta$ functional equation, with one- and two-loop examples}

What is the functional equation obeyed by the local $\beta$ function? \ It
follows simply enough from (\ref{beta}), or else from the definition in
(\ref{beta}) combined with (\ref{Schroeder}). \ Thus,%
\begin{equation}
\frac{\Psi\left(  u\right)  }{\frac{d}{du}\Psi\left(  u\right)  }=\frac
{\Psi\left(  \sigma\left(  u\right)  \right)  }{\frac{d}{du}\Psi\left(
\sigma\left(  u\right)  \right)  }=\frac{1}{\frac{d}{du}\sigma\left(
u\right)  }\frac{\Psi\left(  \sigma\left(  u\right)  \right)  }{\Psi^{\prime
}\left(  \sigma\left(  u\right)  \right)  }\ .
\end{equation}
That is to say (cf. Eq (6) in \cite{CV}),%
\begin{equation}
\beta\left(  \sigma\left(  u\right)  \right)  =\frac{d\sigma\left(  u\right)
}{du}~\beta\left(  u\right)  \ . \label{BetaFcnlEqn}%
\end{equation}
This is just the flow of $\sigma\left(  u\right)  $ along the characteristics
defined by (\ref{characteristics}). \ Note that all \emph{explicit} reference
to the eigenvalue $\lambda$ has dropped out of this equation, although it is
still possible for $\lambda$ dependence to be induced through implicit
$\lambda$ dependence in $\sigma\left(  u\right)  $, and therefore $\lambda$
dependence is implicitly understood for $\beta$ as well. \ Also note that
(\ref{BetaFcnlEqn}) alone does not determine the overall normalization of
$\beta$. \ This normalization is determined by (\ref{beta}) and
(\ref{Schroeder}), and it also introduces $\lambda$ dependence.

As an example, again take $\beta\left(  u\right)  =cu^{2}$, the one-loop
result. \ For $t=1$, we have $\sigma\left(  u\right)  =\frac{u}{1-cu}$, hence
\begin{equation}
\frac{d\sigma\left(  u\right)  }{du}=\frac{1}{\left(  1-cu\right)  ^{2}}\ .
\end{equation}
But we also have
\begin{equation}
\beta\left(  \sigma\left(  u\right)  \right)  =c\sigma^{2}\left(  u\right)
=\frac{cu^{2}}{\left(  1-cu\right)  ^{2}}=\frac{1}{\left(  1-cu\right)  ^{2}%
}~\beta\left(  u\right)  \ .
\end{equation}
So, (\ref{BetaFcnlEqn}) holds --- the most important point to take away from
this example being that a Taylor series solution about $u=0$ \emph{can}
\emph{work} for the functional equation obeyed by $\beta\left(  u\right)  $,
even though it does \emph{not} work for $\Psi\left(  u\right)  $. \ It is of
course redundant to do so, but we check this, for $\beta\left(  u\right)
\equiv au+\gamma u^{2}+bu^{3}$. \ Then $\sigma\left(  u\right)  =\frac
{u}{1-cu}$ gives $\beta\left(  \sigma\left(  u\right)  \right)  -\left(
\frac{d\sigma\left(  u\right)  }{du}\right)  \beta\left(  u\right)
=\frac{cu^{2}}{\left(  cu-1\right)  ^{3}}\left(  -bu^{2}-acu+a\right)  $.
\ For this to vanish, it is necessary and sufficient that both $a=0$ and
$b=0$, while $\gamma$ is undetermined.

\subsubsection{Two-loop infrared fixed point}

For another, more interesting example, consider the two-loop perturbative
approximation to $\beta$ for a model with trivial UV \emph{and} nontrivial IR
fixed points. \ This example nicely illustrates how the normalization of
$\beta$ is determined in the functional approach. \ We may sweep various model
dependent factors into the definition of the coupling, $g$, and the scale of
$t$ to write:%
\begin{equation}
\frac{dg}{dt}=\frac{1}{2}~g^{3}\left(  1-g^{2}\right)  \ . \label{2LoopIR}%
\end{equation}
Changing variable to
\begin{equation}
y=\frac{1}{g^{2}}-1\ ,
\end{equation}
(\ref{2LoopIR}) becomes%
\begin{equation}
\frac{dy}{dt}=\frac{-y}{1+y}\ , \label{2LoopIRAgain}%
\end{equation}
with solution \cite{GGK}%
\begin{equation}
y\left(  t\right)  =\operatorname{LambertW}\left(  y_{0}e^{y_{0}-t}\right)
\ , \label{2LoopSolution}%
\end{equation}
where $\operatorname{LambertW}$\ is the inverse function for $x\exp x$. \ This
is manifestly in FC form (\ref{global}), with $\lambda=1/e$ and $\Psi\left(
y\right)  =y\exp y$, where $\Psi$ satisfies the functional equation%
\begin{equation}
\frac{1}{e}~\Psi\left(  y\right)  =\Psi\left(  \operatorname{LambertW}\left(
\frac{1}{e}~y\exp\left(  y\right)  \right)  \right)  \ . \label{IRSchroeder}%
\end{equation}
From this, we immediately read-off the step-scaling function in terms of $y$.
\ Switching back to the original coupling $g$ the solution
(\ref{2LoopSolution}) gives
\begin{equation}
g^{2}\left(  t\right)  =\frac{1}{1+\operatorname{LambertW}\left(  \left(
\frac{1}{g_{0}^{2}}-1\right)  e^{-t-1+1/g_{0}^{2}}\right)  }\ .
\end{equation}
A typical trajectory is shown here.%
%TCIMACRO{\FRAME{dtbpFU}{4.5405in}{3.0246in}{0pt}{\Qcb{A 2-loop trajectory with
%$g\left(  0\right)  =1/2$.}}{}{schroederrenormprd__1.eps}%
%{\special{ language "Scientific Word";  type "GRAPHIC";
%maintain-aspect-ratio TRUE;  display "USEDEF";  valid_file "F";
%width 4.5405in;  height 3.0246in;  depth 0pt;  original-width 4.5981in;
%original-height 3.0539in;  cropleft "0";  croptop "1";  cropright "1";
%cropbottom "0";
%filename 'SchroederRenormPRD__1.eps';file-properties "XNPEU";}}}%
%BeginExpansion
\begin{center}
\includegraphics[
height=3.0246in,
width=4.5405in
]%
{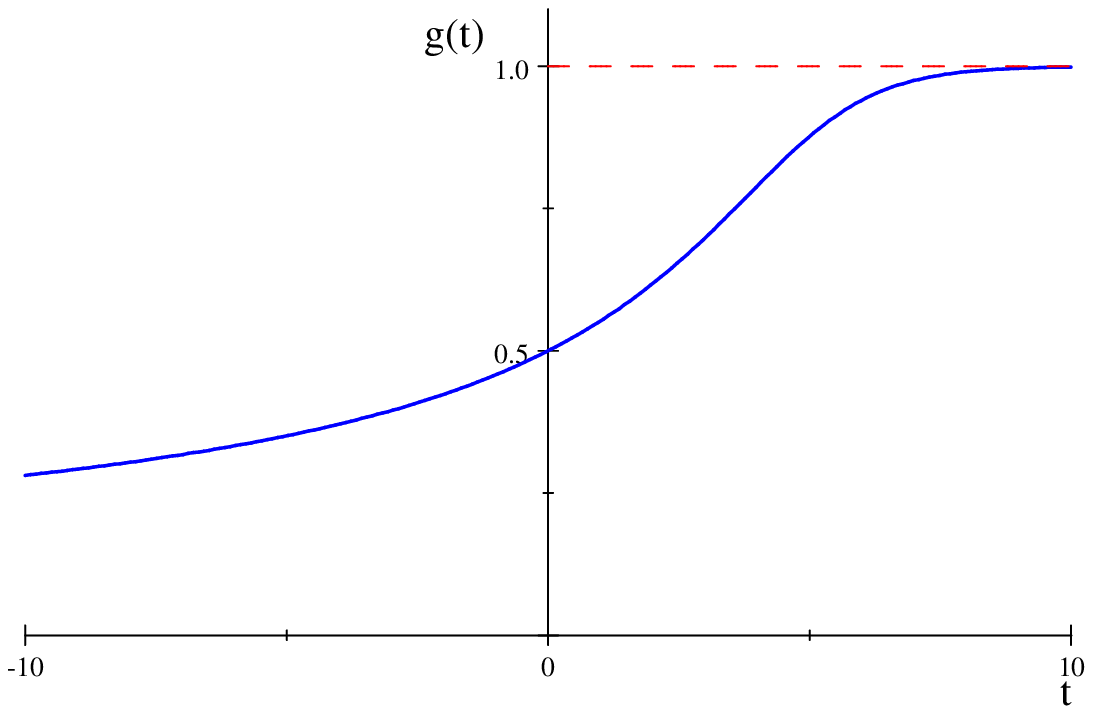}%
\\
A 2-loop trajectory with $g\left(  0\right)  =1/2$.
\end{center}
%EndExpansion

In this example, $\Psi$ \emph{can} be obtained by Taylor series solution of
the functional equation, (\ref{IRSchroeder}), only it is a Taylor series about
the \emph{non-trivial} fixed point at $g=1$, i.e. $y=0$. \ Nonetheless, for
this two-loop example, one may forego the series solution of
(\ref{IRSchroeder}) and just solve it by inspection, upon noting that
$\operatorname{LambertW}^{-1}\left(  z\right)  =ze^{z}$.

For this same example, the $\beta$ function can be obtained \emph{from the
step-scaling function} by series solution about \emph{either} fixed point,
$g=0$ or $g=1$. \ In terms of the variable $y$ with $\sigma\left(  y\right)
=\left.  y\left(  t\right)  \right\vert _{t=1}$,\ we have
\begin{equation}
\sigma\left(  y\right)  =\operatorname{LambertW}\left(  \frac{1}{e}%
~y\exp\left(  y\right)  \right)  \ ,\ \ \ \frac{d\sigma\left(  y\right)  }%
{dy}=\frac{1+y}{y}~\frac{\operatorname{LambertW}\left(  ye^{-1+y}\right)
}{\left(  1+\operatorname{LambertW}\left(  ye^{-1+y}\right)  \right)  }\ .
\end{equation}
This leads to a typical plot of $\sigma$ near the fixed point.%
%TCIMACRO{\FRAME{dtbpFU}{4.5405in}{3.0246in}{0pt}{\Qcb{Two-loop IR fixed point
%exhibited by $\sigma\left(  y\right)  $ (solid red) and $\sigma^{-1}\left(
%y\right)  $ (dashed red) versus $y$. \ Light gray curves are the functional
%square roots of $\sigma$ and $\sigma^{-1}$.}}{}{schroederrenormprd__2.eps}%
%{\special{ language "Scientific Word";  type "GRAPHIC";
%maintain-aspect-ratio TRUE;  display "USEDEF";  valid_file "F";
%width 4.5405in;  height 3.0246in;  depth 0pt;  original-width 4.5981in;
%original-height 3.0539in;  cropleft "0";  croptop "1";  cropright "1";
%cropbottom "0";
%filename 'SchroederRenormPRD__2.eps';file-properties "XNPEU";}}}%
%BeginExpansion
\begin{center}
\includegraphics[
height=3.0246in,
width=4.5405in
]%
{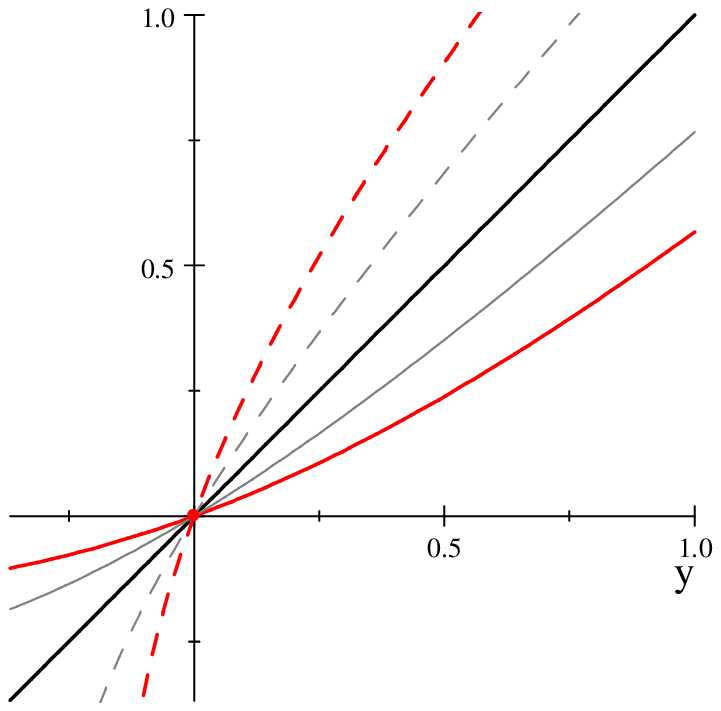}%
\\
Two-loop IR fixed point exhibited by $\sigma\left(  y\right)  $ (solid red)
and $\sigma^{-1}\left(  y\right)  $ (dashed red) versus $y$. \ Light gray
curves are the functional square roots of $\sigma$ and $\sigma^{-1}$.
\end{center}
%EndExpansion

Of course, with all the additional information about the actual trajectory
implicitly built into this closed-form expression for $\sigma$, we may also
forego an actual series solution of (\ref{BetaFcnlEqn}) and solve it too by
inspection. \ Explicitly writing out the functional equation as
\begin{equation}
\beta\left(  \operatorname{LambertW}\left(  ye^{-1+y}\right)  \right)
=\allowbreak\frac{\operatorname{LambertW}\left(  ye^{-1+y}\right)  }{\left(
1+\operatorname{LambertW}\left(  ye^{-1+y}\right)  \right)  }~\frac{1+y}%
{y}~\beta\left(  y\right)  \ ,
\end{equation}
a solution is obvious, namely,\ $\beta\left(  y\right)  \propto\frac{y}{1+y}$.
\ The constant of proportionality is then given by $\ln\lambda=\ln\left(
1/e\right)  =-1$, as in (\ref{beta}), and thus (\ref{2LoopIRAgain}) is recovered.

\subsubsection{M\"{o}bius transformation form}

This is a simple Pad\'{e} approximant \cite{EJJKS}, sometimes known as an
\textquotedblleft exact\textquotedblright\ $\beta$ function\ \cite{R}, and is
but a minor variation on the previous two-loop example. \ For constants
$\mathrm{\alpha}$, $\mathrm{\beta}$, $\mathrm{\gamma}$, and $\mathrm{\delta}$,
consider%
\begin{equation}
\frac{dg}{dt}=\frac{1}{2}~g^{3}~\frac{\mathrm{\alpha}+\mathrm{\beta}g^{2}%
}{\mathrm{\gamma}+\mathrm{\delta}g^{2}}\ .
\end{equation}
Upon changing variables to%
\begin{equation}
y=\frac{\mathrm{\gamma}\left(  \mathrm{\beta}+\mathrm{\alpha}/g^{2}\right)
}{\mathrm{\alpha\delta}-\mathrm{\beta\gamma}}\ ,
\end{equation}
the equation becomes just like (\ref{2LoopIRAgain}) with only a change of
scale for $t$:%
\begin{equation}
\frac{dy}{dt}=\frac{-\mathrm{\alpha}^{2}}{\mathrm{\alpha\delta}-\mathrm{\beta
\gamma}}\frac{y}{1+y}\ .
\end{equation}
Thus the solution is%
\begin{equation}
y\left(  t\right)  =\operatorname{LambertW}\left(  y_{0}e^{y_{0}%
-\frac{\mathrm{\alpha}^{2}}{\mathrm{\alpha\delta}-\mathrm{\beta\gamma}}%
t}\right)  \ .\nonumber
\end{equation}
This is again of FC form with eigenvalue $\lambda=e^{\frac{-\mathrm{\alpha
}^{2}}{\mathrm{\alpha\delta}-\mathrm{\beta\gamma}}}$. \ In terms of the
original variable the RG trajectory is given by%
\begin{equation}
g^{2}\left(  t\right)  =\frac{-\mathrm{\alpha}/\mathrm{\beta}}{1-\left(
\frac{\mathrm{\alpha\delta}-\mathrm{\beta\gamma}}{\mathrm{\beta\gamma}%
}\right)  \operatorname{LambertW}\left(  \frac{\mathrm{\gamma}}{\mathrm{\alpha
\delta}-\mathrm{\beta\gamma}}\left(  \mathrm{\beta}+\dfrac{\mathrm{\alpha}%
}{g^{2}}\right)  \exp\left(  \frac{\mathrm{\gamma}}{\mathrm{\alpha\delta
}-\mathrm{\beta\gamma}}\left(  \mathrm{\beta}+\dfrac{\mathrm{\alpha}}{g^{2}%
}\right)  -\frac{1}{\mathrm{\alpha\delta}-\mathrm{\beta\gamma}}\mathrm{\alpha
}^{2}t\right)  \right)  }\ .
\end{equation}

\subsubsection{One-loop geometromorphosis}

This describes the renormalization flow of geometry from a flat manifold in
the UV towards a fixed, nontrivial manifold in the IR. \ The trajectories
describe evolution of matrices,%
\begin{equation}
\frac{dg_{ab}\left(  t\right)  }{dt}=\beta_{ab}\left(  t\right)
\ ,\ \ \ g_{ab}\left(  t\right)  =g_{ab}\left(  0\right)  +\int_{0}^{t}%
d\tau~\beta_{ab}\left(  \tau\right)  \ .
\end{equation}
For RG flow of this type, we would expect a tensor version of the Schr\"{o}der
equation to be operative,%
\begin{equation}
\Psi_{ab}\left(  \mathbf{g}\left(  1\right)  \right)  =\Lambda_{ac}\Psi
_{cb}\left(  \mathbf{g}\left(  0\right)  \right)  \ ,
\end{equation}
where $g_{ab}$ could include the torsion potential as well as the metric (in
which case $g_{ab}\neq g_{ba}$) and where the step-scaling function is now a
RG transported $g_{ab}$,
\begin{equation}
g_{ab}\left(  1\right)  =\Lambda_{ac}g_{cb}\left(  0\right)  \text{ ,
\ \ }\Lambda_{ac}=\delta_{ac}+\left(  \int_{0}^{1}d\tau~\beta_{ad}\left(
\tau\right)  \right)  g_{db}^{-1}\left(  0\right)  \ .
\end{equation}

To simplify the discussion, and to be explicit, consider the 3-sphere $\sigma
$-model with torsion \cite{BCZ}, with $\mathfrak{S}$ proportional to the
square of the radius of $S_{3}$. \ In this case, the one-loop renormalization
of the metric boils down to just a change of $\mathfrak{S}$ with length
scale:
\begin{equation}
\frac{d\mathfrak{S}}{dt}=\frac{1}{2}\left(  \frac{1}{\mathfrak{S}^{2}%
}-1\right)  \ .
\end{equation}
The solution of this one-loop evolution equation is given implicitly by%
\begin{equation}
\frac{\mathfrak{S}\left(  t\right)  +1}{\mathfrak{S}\left(  t\right)
-1}~e^{-2\mathfrak{S}\left(  t\right)  }=e^{t}~\frac{\mathfrak{S}%
+1}{\mathfrak{S}-1}~e^{-2\mathfrak{S}}\ .
\end{equation}
That is to say, (\ref{Psi(u(t))}) and (\ref{Schroeder}) have eigenvalue
$\lambda=e$, with explicit Schr\"{o}der function
\begin{equation}
\Psi\left(  \mathfrak{S}\right)  =\frac{\mathfrak{S}+1}{\mathfrak{S}%
-1}~e^{-2\mathfrak{S}}\ ,
\end{equation}
while $\Psi^{-1}$ is only implicit. \ Note that $\Psi\left(  \mathfrak{S}%
\right)  >$ $0$ when $\mathfrak{S}>1$. \ The 3-sphere squared-radius RG
evolution is then given in FC form by
\begin{equation}
\mathfrak{S}\left(  t\right)  =\Psi^{-1}\left(  e^{t}\Psi\left(
\mathfrak{S}\right)  \right)  \ ,
\end{equation}
This defines implicitly the step-scaling function $\sigma\left(
\mathfrak{S}\right)  $, as $\mathfrak{S}\left(  1\right)  $, say, shown here.
\ The fixed point $\sigma\left(  \mathfrak{S}_{\ast}\right)  =\mathfrak{S}%
_{\ast}=1$ is centered in the small black circle.%
%TCIMACRO{\FRAME{dtbpFU}{4.5777in}{3.5142in}{0pt}{\Qcb{One-loop
%\textquotedblleft geometrostatic\textquotedblright\ IR fixed point exhibited
%by $\sigma\left(  \QTR{frak}{S}\right)  $ (solid red) and $\sigma^{-1}\left(
%\QTR{frak}{S}\right)  $ (dashed red) versus $\QTR{frak}{S}$, and a few other
%fixed $t$ slices of the $\sigma_{t}\left(  \QTR{frak}{S}\right)  $ surface
%(thin gray curves).}}{}{schroederrenormprd__3.eps}%
%{\special{ language "Scientific Word";  type "GRAPHIC";
%maintain-aspect-ratio TRUE;  display "USEDEF";  valid_file "F";
%width 4.5777in;  height 3.5142in;  depth 0pt;  original-width 5.0123in;
%original-height 3.8406in;  cropleft "0";  croptop "1";  cropright "1";
%cropbottom "0";
%filename 'SchroederRenormPRD__3.eps';file-properties "XNPEU";}}}%
%BeginExpansion
\begin{center}
\includegraphics[
height=3.5142in,
width=4.5777in
]%
{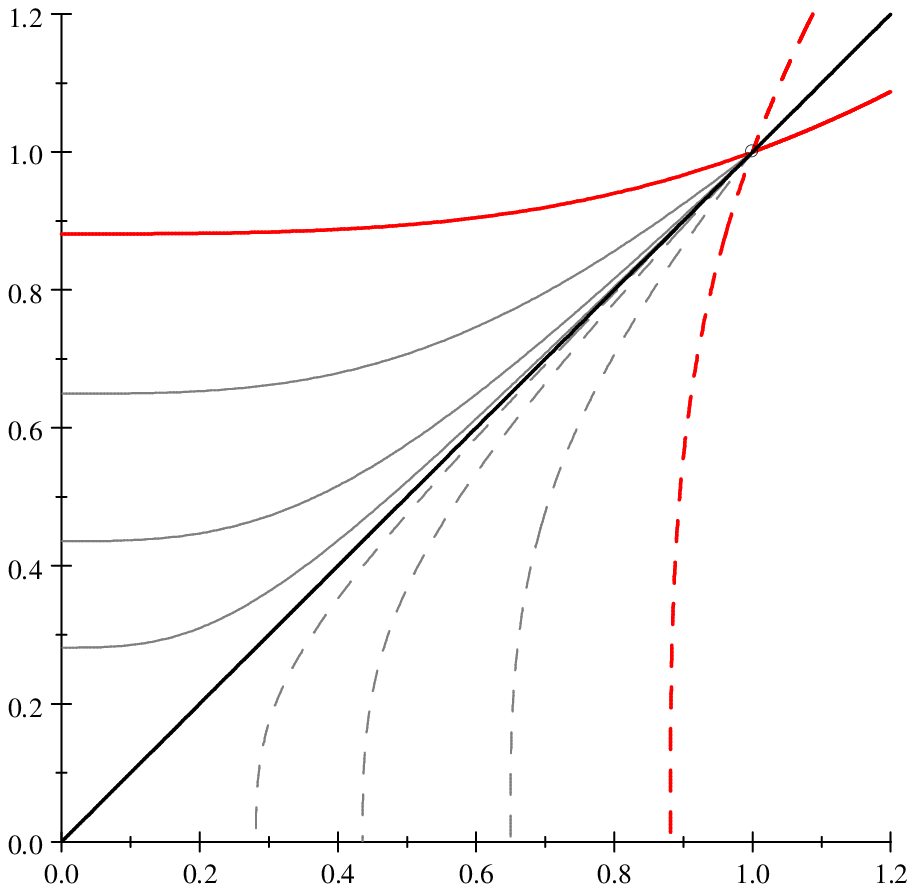}%
\\
One-loop \textquotedblleft geometrostatic\textquotedblright\ IR fixed point
exhibited by $\sigma\left(  \mathfrak{S}\right)  $ (solid red) and
$\sigma^{-1}\left(  \mathfrak{S}\right)  $ (dashed red) versus $\mathfrak{S}$,
and a few other fixed $t$ slices of the $\sigma_{t}\left(  \mathfrak{S}%
\right)  $ surface (thin gray curves).
\end{center}
%EndExpansion

The implicit function $\Psi^{-1}$ has no name, as far as we can tell, although
it might be classified as a generalization of the Lambert function. \ In any
case, we may construct the inverse function $\Psi^{-1}\left(  z\right)  $
through the usual graphical methods, to find two branches.%
%TCIMACRO{\FRAME{dtbpFU}{4.5857in}{3.7901in}{0pt}{\Qcb{Two branches of
%$\Psi^{-1}\left(  z\right)  $, in green, and various approximations (orange
%dashes).}}{}{schroederrenormprd__4.eps}{\special{ language "Scientific Word";
%type "GRAPHIC";  maintain-aspect-ratio TRUE;  display "USEDEF";
%valid_file "F";  width 4.5857in;  height 3.7901in;  depth 0pt;
%original-width 5.0123in;  original-height 4.1378in;  cropleft "0";
%croptop "1";  cropright "1";  cropbottom "0";
%filename 'SchroederRenormPRD__4.eps';file-properties "XNPEU";}}}%
%BeginExpansion
\begin{center}
\includegraphics[
height=3.7901in,
width=4.5857in
]%
{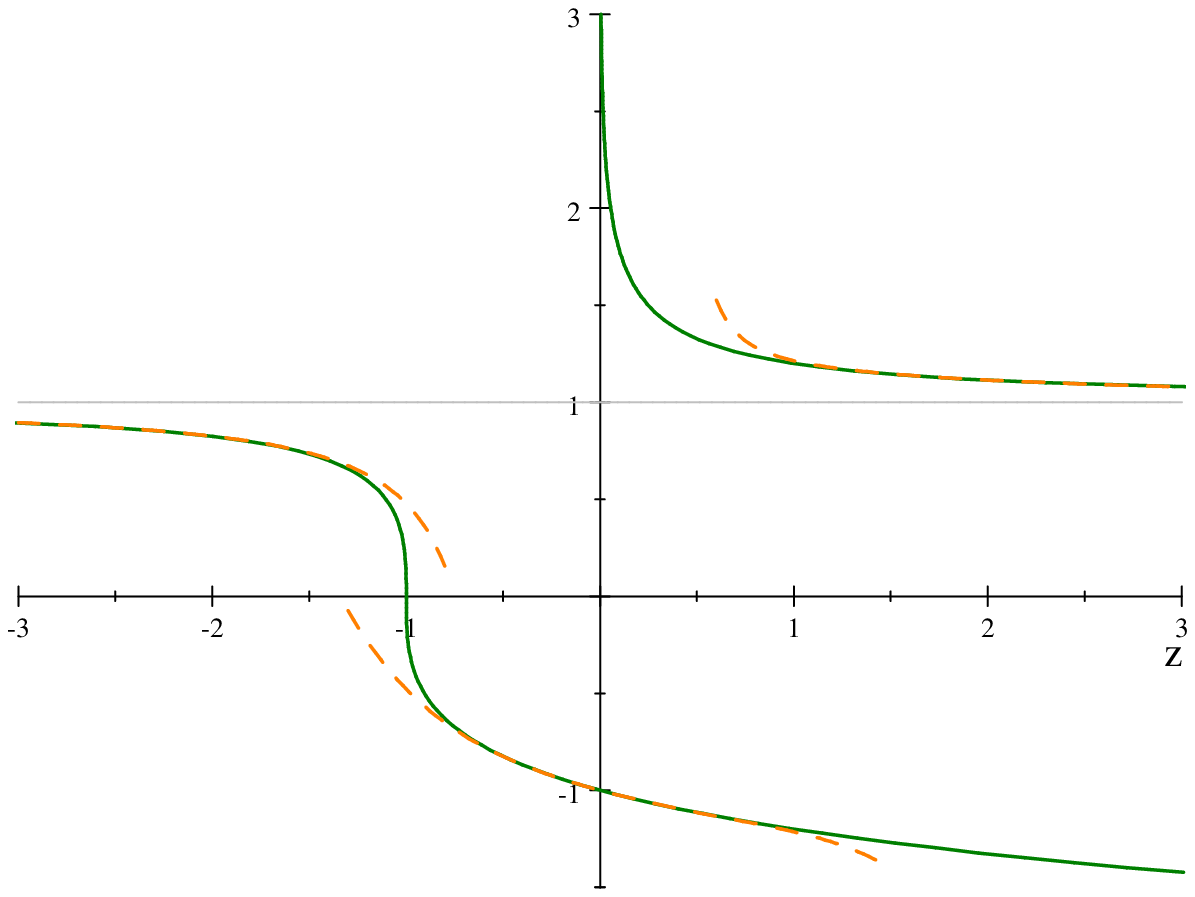}%
\\
Two branches of $\Psi^{-1}\left(  z\right)  $, in green, and various
approximations (orange dashes).
\end{center}
%EndExpansion
The UV fixed point --- a 3-sphere of infinite radius --- corresponds to the
vertical asymptote of the \emph{upper} branch of $\Psi^{-1}$, while the IR
fixed point --- a 3-sphere of unit radius --- corresponds to the horizontal
asymptote (thin light gray line in the Figure) of that upper branch. \ Also as
shown in the Figure, it is straightforward to construct series approximations
around the point $\left(  z,\Psi^{-1}\left(  z\right)  \right)  =\left(
0,-1\right)  $, for the lower branch,%
\begin{equation}
\Psi_{\text{lower branch}}^{-1}\left(  z\right)  =-1-2e^{-2}z+6e^{-4}%
z^{2}-26e^{-6}z^{3}+\frac{2}{3}197e^{-8}z^{4}-722e^{-10}z^{5}+O\left(
z^{6}\right)  \ ,
\end{equation}
as well as around the point(s) $\left(  z,\Psi^{-1}\left(  z\right)  \right)
=\left(  \pm\infty,+1\right)  $, for either the upper or the lower branches,%
\begin{equation}
\Psi^{-1}\left(  z\right)  =1+2e^{-2}\left(  \frac{1}{z}\right)
-6e^{-4}\left(  \frac{1}{z}\right)  ^{2}+26e^{-6}\left(  \frac{1}{z}\right)
^{3}-\frac{2}{3}197e^{-8}\left(  \frac{1}{z}\right)  ^{4}+722e^{-10}\left(
\frac{1}{z}\right)  ^{5}+O\left(  \left(  \frac{1}{z}\right)  ^{6}\right)  \ .
\end{equation}
The second of these series follows from the first through the transformation
$\left(  \Psi,\mathfrak{S}\right)  \rightarrow\left(  \frac{1}{\Psi
},-\mathfrak{S}\right)  $ applied to the equation for $\Psi$. \ These series
results for $\Psi^{-1}$ are representative of situations where simple,
closed-form expressions are not readily available. \ A similar situation often
arises when we have:

\section{A lattice gauge theory model}

In this section we consider a physical illustration of fixed point behavior
drawn from numerical studies of lattice gauge theory \cite{A}. \ \ \ The
$\beta$ and $\sigma$ functions in question are those for a non-abelian gauge
theory with 12 flavors of $su\left(  3\right)  $ color triplets.

\subsection{Parameterizations\ }

It is convenient in lattice gauge theory to approximate the physical coupling
at length scale $\ell$, $u=1/g^{2}\left(  \ell\right)  $, parametrically in
terms of the bare lattice coupling, $s=1/g_{0}^{2}$, as a series \cite{A},
\begin{equation}
u=\frac{1}{g^{2}\left(  \ell\right)  }=s\left(  1-\sum_{j=1}^{n}c_{j}\left(
\ell\right)  \frac{1}{s^{j}}\right)  \ . \label{u}%
\end{equation}
The step-scaling procedure \cite{C,L} then gives the coupling at length $L$ as
a function of $u$. \ Of course, this is also a function of the bare lattice
coupling parameter, $s$, but with a different series expansion, in general,
perhaps even with a different order for the series (depending on the choices
made in the numerical computations).%
\begin{equation}
\sigma\left(  u\right)  =\frac{1}{g^{2}\left(  L\right)  }=s\left(
1-\sum_{j=1}^{N}c_{j}\left(  L\right)  \frac{1}{s^{j}}\right)  \ .
\label{sigma}%
\end{equation}
Note that these series are arranged to have a common zero-coupling limit, as
$s\rightarrow\infty$, with both (\ref{u}) and (\ref{sigma}) becoming the
identity map in that limit.\ \ For this parameterization, the functional
equation (\ref{BetaFcnlEqn}) for $\beta\left(  u\right)  $ may be written as%
\begin{equation}
\beta\left(  \sigma\left(  u\left(  s\right)  \right)  \right)  ~\frac
{du\left(  s\right)  }{ds}=\beta\left(  u\left(  s\right)  \right)
~\frac{d\sigma\left(  u\left(  s\right)  \right)  }{ds}\ ,
\label{BetaFcnlEqnParametric}%
\end{equation}
where%
\begin{equation}
\frac{du\left(  s\right)  }{ds}=1+\sum_{j=1}^{n}\left(  j-1\right)
c_{j}\left(  \ell\right)  \frac{1}{s^{j}}\text{ \ \ and \ \ }\frac
{d\sigma\left(  u\left(  s\right)  \right)  }{ds}=1+\sum_{j=1}^{N}\left(
j-1\right)  c_{j}\left(  L\right)  \frac{1}{s^{j}}\ . \label{du&dsigma}%
\end{equation}
It is sensible from the stand-point of perturbation theory to consider a
similar expansion for $\beta$. \ Thus we write%
\begin{equation}
\beta\left(  u\right)  =\sum_{n\geq0}\frac{b_{n}}{u^{n}}\ .
\end{equation}
Using this series along with (\ref{u}), (\ref{sigma}), and (\ref{du&dsigma}),
and expanding both sides of (\ref{BetaFcnlEqnParametric}) in powers of $1/s$,
we obtain recursion relations for the coefficient ratios $b_{n}/b_{0}$. \ The
$O\left(  1\right)  $ and $O\left(  1/s\right)  $ terms on LHS and RHS of
(\ref{BetaFcnlEqnParametric}) match identically, but the $O\left(
1/s^{n}\right)  $ terms for $n\geq2$ give expressions for $b_{n-1}/b_{0}$ in
terms of the $c_{k\leq n}$. \ For example, we find%
\begin{equation}
\frac{b_{1}}{b_{0}}=\frac{c_{2}\left(  L\right)  -c_{2}\left(  \ell\right)
}{c_{1}\left(  L\right)  -c_{1}\left(  \ell\right)  }\ ,\ \ \ \frac{b_{2}%
}{b_{0}}=\frac{2c_{3}\left(  L\right)  -2c_{3}\left(  \ell\right)  -\left(
c_{1}\left(  L\right)  +c_{1}\left(  \ell\right)  \right)  \left(
c_{2}\left(  L\right)  -c_{2}\left(  \ell\right)  \right)  }{c_{1}\left(
L\right)  -c_{1}\left(  \ell\right)  }\ ,
\end{equation}
etc. \ The overall normalization of $\beta$\ is not determined by
(\ref{BetaFcnlEqnParametric}), of course, and in fact there is no information
in the expansions (\ref{u}), (\ref{sigma}), and (\ref{du&dsigma})\ that allows
determination of $b_{0}$ by Taylor expanding about $s=\infty$. \ (This is
related to the essential singularity in the Schr\"{o}der function solutions at
zero coupling, as mentioned earlier, (\ref{SigmaSeries}) et seq.)\ \ Rather,
we must fix $b_{0}$ by other considerations, the obvious choice being to use
perturbation theory. \ Another possibility is to use lattice data and expand,
not about zero coupling, but about a nontrivial fixed point, if available.
\ The expansions are similar to those we have just given, and are collected
together in the Appendix, \S IX. \ 

\subsection{Numerics}

Considerable effort is needed to properly take the continuum limit where the
lattice spacing $a$ goes to zero. \ However, for purposes of
\emph{illustration} of the various functional methods described here, we will
not concern ourselves with those complications. \ Rather, we will simply take
some of the raw numerical data in \cite{A} for the expansion coefficients
appearing in (\ref{u}) and (\ref{sigma}), and note with amusement that such a
naive \emph{ab initio} computation of the $\beta$ function matches very well
with two-loop perturbation theory, upon overlapping the two results.

We choose to consider the $L=8a$ and $L=16a$ data from \cite{A} for 12 flavors
of $su\left(  3\right)  $ color triplets, a model widely believed to have a
nontrivial IR fixed point near $g^{2}=5$. \ We take the data at face value,
without regard for any statistical or systematic errors, and we use this data
for the parametric definitions of $u$ and $\sigma\left(  u\right)  $.
\begin{align}
u\left(  s\right)   &  =s\left(  1-0.4092\left(  \frac{1}{s}\right)
+0.192\left(  \frac{1}{s}\right)  ^{2}-0.73\left(  \frac{1}{s}\right)
^{3}+0.837\left(  \frac{1}{s}\right)  ^{4}-0.342\left(  \frac{1}{s}\right)
^{5}\right)  \ ,\nonumber\\
\sigma\left(  u\left(  s\right)  \right)   &  =s\left(  1-0.467\left(
\frac{1}{s}\right)  +0.154\left(  \frac{1}{s}\right)  ^{2}-0.164\left(
\frac{1}{s}\right)  ^{3}\right)  \ . \label{Appelquist8&16}%
\end{align}
For illustration purposes, this will suffice; but, hopefully, the procedures
we follow will be useful in future, realistic lattice studies. \ 

In any case, we plot $\sigma\left(  u\right)  $, parametrically, versus $u$ to
display the $\sigma\left(  u_{\ast}\right)  =u_{\ast}=1/g_{\ast}^{2}$ fixed
point encoded in the data, as given numerically by:%
\begin{equation}
g_{\ast}^{2}=\frac{1}{0.180}=5.\,56\ \ \ \text{at \ }%
s=0.739\ ,\ \ \ \text{i.e. \ }g_{0\ast}^{2}=\frac{1}{0.739}=1.\,35\ .
\label{LatticeIRFP}%
\end{equation}
The two series (\ref{Appelquist8&16}) were \emph{constructed} to become
parallel curves in the zero coupling limit, as $s$ or $u\rightarrow\infty$.
\ However, the curves have significantly different approaches to the
nontrivial fixed point, hence its existence, as is evident in the Figure.
\ The fixed point $u_{\ast}=0.180$ is centered in the small black circle.%
%TCIMACRO{\FRAME{dtbpFU}{5.1862in}{4.2433in}{0pt}{\Qcb{$\sigma\left(  u\right)
%$ (solid red) and $\sigma^{-1}\left(  u\right)  $ (dashed red) versus $u$, and
%a few other fixed $t$ slices of the $\sigma_{t}\left(  u\right)  $ surface
%(thin gray curves).}}{}{schroederrenormprd__5.eps}%
%{\special{ language "Scientific Word";  type "GRAPHIC";
%maintain-aspect-ratio TRUE;  display "USEDEF";  valid_file "F";
%width 5.1862in;  height 4.2433in;  depth 0pt;  original-width 5.2563in;
%original-height 4.2957in;  cropleft "0";  croptop "1";  cropright "1";
%cropbottom "0";
%filename 'SchroederRenormPRD__5.eps';file-properties "XNPEU";}}}%
%BeginExpansion
\begin{center}
\includegraphics[
height=4.2433in,
width=5.1862in
]%
{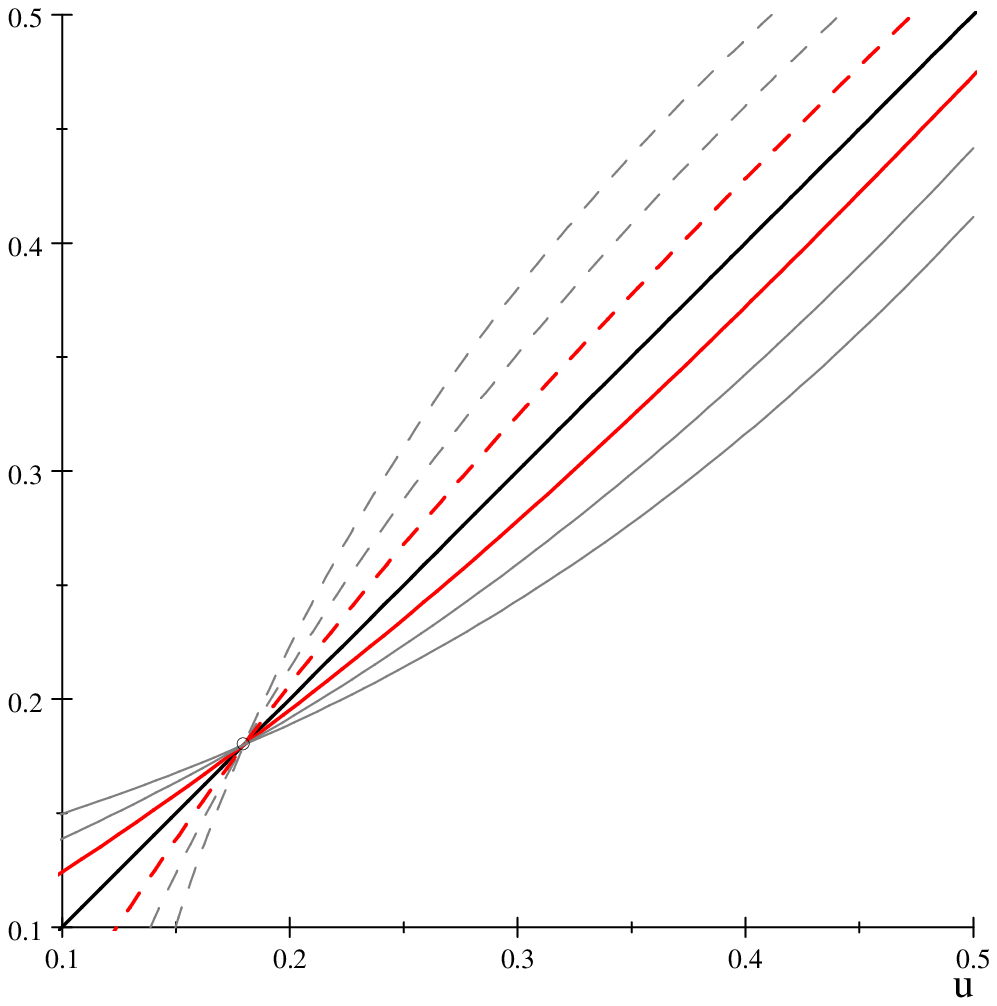}%
\\
$\sigma\left(  u\right)  $ (solid red) and $\sigma^{-1}\left(  u\right)  $
(dashed red) versus $u$, and a few other fixed $t$ slices of the $\sigma
_{t}\left(  u\right)  $ surface (thin gray curves).
\end{center}
%EndExpansion

Now, if we use the $s$-parameterization formalism discussed above, we obtain
for example:%
\begin{equation}
\frac{b_{1}}{b_{0}}=0.657\ ,\ \ \ \frac{b_{2}}{b_{0}}=-20.2\ . \label{rawBs}%
\end{equation}
This certainly \emph{looks} useless (i.e. like a divergent series for
$g^{2}\gtrapprox0.03$) and is very much in disagreement with perturbation
theory, as given to 2-loops for the 12 flavor $su\left(  3\right)  $ model by
\begin{equation}
\frac{d}{dt}u=-2\left(  \frac{3}{\left(  4\pi\right)  ^{2}}-\frac{50}{\left(
4\pi\right)  ^{4}}\frac{1}{u}\right)  \ , \label{PertTheo}%
\end{equation}
such that $b_{1}/b_{0}=-\frac{50}{3\left(  4\pi\right)  ^{2}}=-0.106$. \ But
if nothing else, (\ref{rawBs}) confirms that the data in (\ref{Appelquist8&16}%
) certainly did not come just from transcribing perturbation theory.

On the other hand, if we reparameterize the data around the IR fixed point,
defining
\begin{equation}
\frac{1}{s}=\left(  1+r\right)  g_{0\ast}^{2}\ ,
\end{equation}
and expand various quantitites to and including $O\left(  r^{4}\right)  $, as
described in detail in an Appendix, we find%
\begin{align}
u\left(  r\right)   &  =0.180-1.\,52r-1.\,258r^{2}-3.\,25r^{3}-0.408r^{4}%
-0.739r^{5}+O\left(  r^{6}\right)  \ ,\nonumber\\
\sigma\left(  u\left(  r\right)  \right)   &  =0.180-1.\,13r+0.439r^{2}%
-0.739r^{3}+0.739r^{4}+O\left(  r^{5}\right)  \ .
\end{align}
Admittedly, the second of these $r$-series is less firm, since we only have
$\sigma\left(  u\left(  s\right)  \right)  $ to $O\left(  \left(  \frac{1}%
{s}\right)  ^{3}\right)  $, although the difference will turn out to be slight
over the region where the series is reliable. \ We describe below the changes
encountered from truncating the $\sigma\left(  u\left(  r\right)  \right)  $
series at $O\left(  r^{3}\right)  $ versus $O\left(  r^{4}\right)  $.

Using again the functional formalism to determine the $\beta$ function from
$\sigma$, only now as implemented in terms of the expansion about the
nontrivial fixed point (for details, see the Appendix), we find the more
sensible expression%
\begin{equation}
\beta\left(  u\right)  =\beta\left(  u_{\ast}+w\right)  =\mathfrak{b}%
_{1}\times\left(  w-3.14w^{2}+1.\,86w^{3}+4.\,02w^{4}\right)
\ ,\ \ \ \text{where \ \ }\mathfrak{b}_{1}=\ln\left(  0.745\right)  =-0.294\ .
\label{LatticeBeta}%
\end{equation}
This is the best one can do given only the data in (\ref{Appelquist8&16}).
\ So, small, positive initial $w$ will decrease as $t$ increases. \ That is to
say, $u>u_{\ast}$ will decrease down to $u_{\ast}$ as $t$ increases. \ Or,
since $u=1/g^{2}$, $g^{2}<g_{\ast}^{2}$ will increase up to $g_{\ast}^{2}$ as
$t$ increases. \ Increasing $t$ corresponds to increasing length scale, so the
fixed point is indeed an IR one. \ This is clear from computing the
trajectories that follow from (\ref{LatticeBeta}). \ Here is an example of
$g^{2}\left(  t\right)  $, with $g^{2}\left(  0\right)  =3$.%
%TCIMACRO{\FRAME{dtbpFU}{4.5405in}{3.0246in}{0pt}{\Qcb{Lattice (solid blue) and
%2-loop (dashed blue) $g^{2}\left(  t\right)  $ trajectories, with
%$g^{2}\left(  0\right)  =3$.}}{}{schroederrenormprd__6.eps}%
%{\special{ language "Scientific Word";  type "GRAPHIC";
%maintain-aspect-ratio TRUE;  display "USEDEF";  valid_file "F";
%width 4.5405in;  height 3.0246in;  depth 0pt;  original-width 4.5981in;
%original-height 3.0539in;  cropleft "0";  croptop "1";  cropright "1";
%cropbottom "0";
%filename 'SchroederRenormPRD__6.eps';file-properties "XNPEU";}}}%
%BeginExpansion
\begin{center}
\includegraphics[
height=3.0246in,
width=4.5405in
]%
{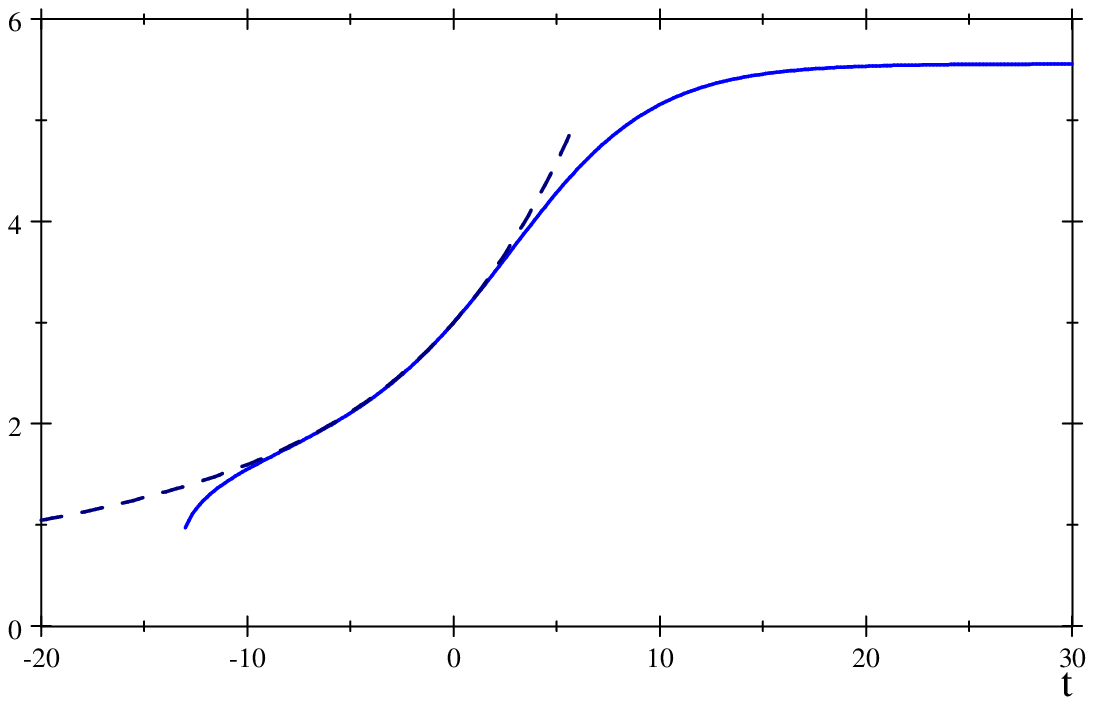}%
\\
Lattice (solid blue) and 2-loop (dashed blue) $g^{2}\left(  t\right)  $
trajectories, with $g^{2}\left(  0\right)  =3$.
\end{center}
%EndExpansion
In terms of both slope and curvature (cf. cubic splines), the trajectory joins
very smoothly with that obtained from two-loop perturbation theory for 12
flavors of quark color triplets.

Further comparison can be made to trajectories determined by other
approximations. \ It is more direct, however, to just compare $\beta$
functions, now that we have in hand (\ref{LatticeBeta}). \ In addition to the
two-loop approximation, which is unchanged by analytic redefinitions of the
coupling, we also compare to three-loop (minimally subtracted, as well as that
based on the Schr\"{o}dinger --- not Schr\"{o}der --- functional) and to
four-loop results from perturbation theory (see \cite{A}\ for the literature
dealing with these approximations). \ Note how well the two-loop result joins
smoothly to our naive, but ab initio determination of $\beta$ using the
lattice data and functional methods, at $g^{2}=3$. \ We stress that there have
been \emph{no} adjustments in the normalizations to facilitate this match-up.
\ The other, higher-loop approximations do not fare nearly as well in terms of
matching-up with the lattice $\beta$, although they do give estimates of
$g_{\ast}^{2}$ more or less in line with (\ref{LatticeIRFP}).%
%TCIMACRO{\FRAME{dtbpFU}{4.5405in}{3.0246in}{0pt}{\Qcb{Various $\beta\left(
%g^{2}\right)  $ functions: \ 2-loop (black), 3-loop MS (green), 3-loop SFR
%(orange), 4-loop MS (blue), lattice (red).}}{}{schroederrenormprd__7.eps}%
%{\special{ language "Scientific Word";  type "GRAPHIC";
%maintain-aspect-ratio TRUE;  display "USEDEF";  valid_file "F";
%width 4.5405in;  height 3.0246in;  depth 0pt;  original-width 4.5981in;
%original-height 3.0539in;  cropleft "0";  croptop "1";  cropright "1";
%cropbottom "0";
%filename 'SchroederRenormPRD__7.eps';file-properties "XNPEU";}}}%
%BeginExpansion
\begin{center}
\includegraphics[
height=3.0246in,
width=4.5405in
]%
{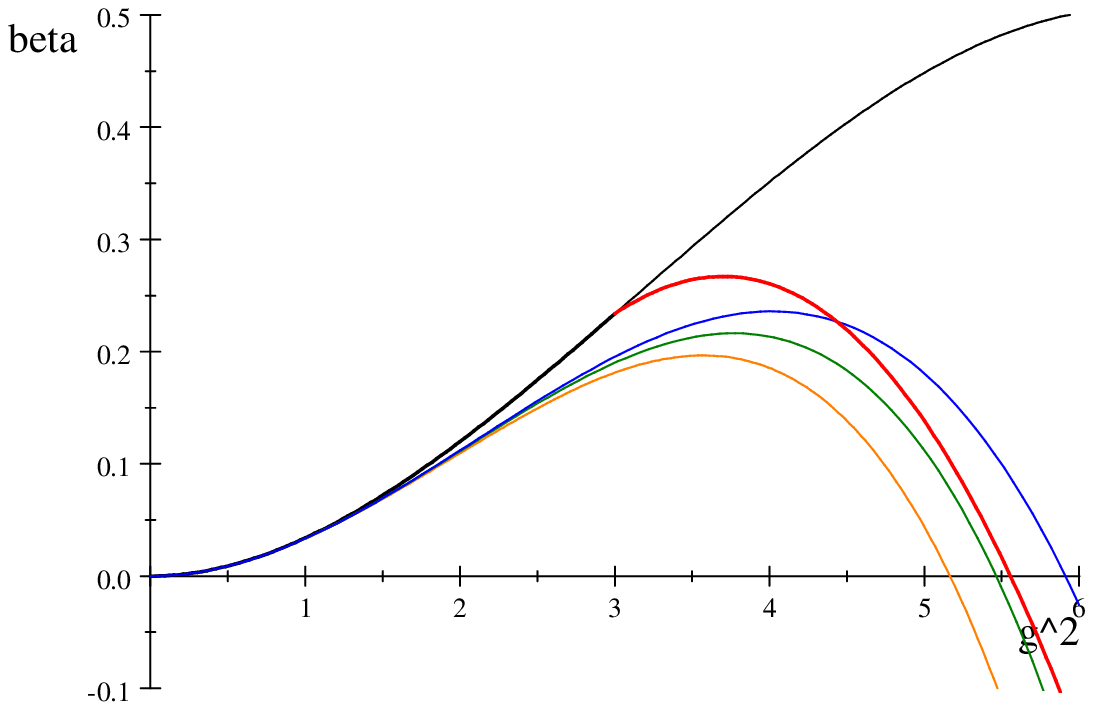}%
\\
Various $\beta\left(  g^{2}\right)  $ functions: \ 2-loop (black), 3-loop MS
(green), 3-loop SFR (orange), 4-loop MS (blue), lattice (red).
\end{center}
%EndExpansion

Finally, we contrast the differences resulting from truncating the
$\sigma\left(  u\left(  r\right)  \right)  $ series at $O\left(  r^{3}\right)
$ versus $O\left(  r^{4}\right)  $. \ The resulting changes in $\beta\left(
g^{2}\right)  $, as obtained from the step-scaling function, are shown below.
\ The difference is slight near $g^{2}=3$, but becomes a sizeable disagreement
for $g^{2}\lessapprox2.5$, at which point the disparity is comparable to that
between (\ref{LatticeBeta}) and the higher-loop approximations at $g^{2}=3$.
\ In fact, the $O\left(  r^{3}\right)  $\ truncation even gives another,
spurious(!) nontrivial fixed point at $g^{2}\approx1.7$. \ Simply put, for
$g^{2}\lessapprox2.5$ the $O\left(  w^{4}\right)  $\ series in
(\ref{LatticeBeta}) is woefully inadequate. \ A better approximation to the
solution of the $\beta$ functional equation must be used for smaller $g^{2}$.%
%TCIMACRO{\FRAME{dtbpFU}{4.5405in}{3.0246in}{0pt}{\Qcb{Two-loop (black) and
%lattice (red) $\beta$s, the latter to both cubic (dashed) and quartic order.}%
%}{}{schroederrenormprd__8.eps}{\special{ language "Scientific Word";
%type "GRAPHIC";  maintain-aspect-ratio TRUE;  display "USEDEF";
%valid_file "F";  width 4.5405in;  height 3.0246in;  depth 0pt;
%original-width 4.5981in;  original-height 3.0539in;  cropleft "0";
%croptop "1";  cropright "1";  cropbottom "0";
%filename 'SchroederRenormPRD__8.eps';file-properties "XNPEU";}}}%
%BeginExpansion
\begin{center}
\includegraphics[
height=3.0246in,
width=4.5405in
]%
{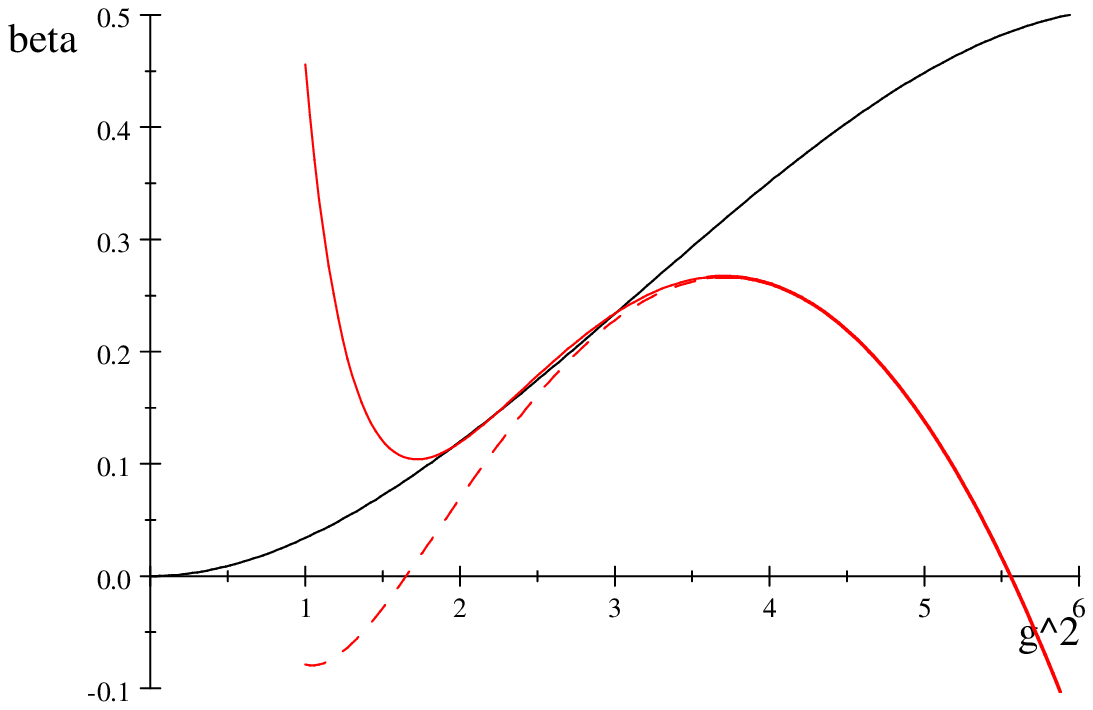}%
\\
Two-loop (black) and lattice (red) $\beta$s, the latter to both cubic (dashed)
and quartic order.
\end{center}
%EndExpansion

While some of these numerical coincidences may very well be little more than
artefacts of the data selected, it is of interest to see if this match-up
between perturbative results, and those obtained for $\beta$ by functional
methods, persists given more accurate lattice data and a thorough analysis of
numerical errors. \ The above discussion is an illustrative application of the
functional methods introduced here, rather than an endorsement of specific results.

\section{A model with a limit cycle}

In this section we illustrate elementary limit cycle behavior in an exactly
solvable lattice model, albeit one-dimensional, as obtained in \cite{LeC} by
an extension of the standard BCS Hamiltonian to include a term that breaks
time-reversal invariance. \ A model with similar RG structure was studied in
\cite{GW}. \ As before, we emphasize the connections between discrete and
continuous scaling, and the underlying functional relationships.

The physics of the model is explained in the extensive work of LeClair et al.
and only the key mathematical features will only be summarized here. \ The
Hamiltonian for the model is%
\begin{equation}
H=\sum_{j}\varepsilon_{j}b_{j}^{\dag}b_{j}+\sum_{j,k}V_{jk}b_{j}^{\dag}%
b_{k}\ ,\ \ \ \ \ V_{jk}=\left\{
\begin{array}
[c]{ccc}%
\left(  g+ih\right)  \epsilon & \text{if} & \varepsilon_{j}>\varepsilon_{k}\\
g\epsilon & \text{if} & \varepsilon_{j}=\varepsilon_{k}\\
\left(  g-ih\right)  \epsilon & \text{if} & \varepsilon_{j}<\varepsilon_{k}%
\end{array}
\right.  \ , \label{ModifiedBCS}%
\end{equation}
where $b_{j}$\ and $b_{j}^{\dag}$\ denote the usual Cooper-pair annihilation
and creation operators, and where $\epsilon=\frac{1}{2}\left(  \varepsilon
_{j+1}-\varepsilon_{j}\right)  $ is the single-particle level spacing. \ 

For a large number of system sites, the renormalization of the dimensionless
couplings $g$ and $h$ under a change in system size $L$ is given by%
\begin{equation}
\frac{dg}{d\ln L}=g^{2}+h^{2}\ ,\ \ \ h=\text{constant\ ,}%
\end{equation}
with $h$ the time-reversal breaking parameter. \ Assuming $h\neq0$, we change
variables to $u=g/h$ and $t=h\ln L$. \ Then
\begin{equation}
\frac{du}{dt}=1+u^{2}\ , \label{ModBCSBeta}%
\end{equation}
and direct integration gives%
\begin{equation}
u\left(  t\right)  =\tan\left(  t+\arctan u_{0}\right)  \ . \label{Direct}%
\end{equation}
Thus the physics of the model repeats itself cyclically as the logarithm of
the system size is changed.

On the other hand, the functional conjugacy formalism gives
\begin{align}
\Psi\left(  u\right)   &  =\exp\left(  \int^{u}\frac{dw}{1+w^{2}}\right)
=\exp\left(  \arctan u\right)  \ ,\ \ \ \Psi^{-1}\left(  u\right)
=\tan\left(  \ln u\right) \\
u\left(  t\right)   &  =\Psi^{-1}\left(  \lambda^{t}\Psi\left(  u_{0}\right)
\right)  =\tan\left(  t\ln\lambda+\arctan u_{0}\right)
\label{ModBCSTrajectory}%
\end{align}
Comparing the last expression to (\ref{Direct}) we see that the Schr\"{o}der
eigenvalue for a unit step in $t$ is $\lambda=e$. \ 

In general, the step-scaling function corresponding to $t$ step size $\Delta$
is
\begin{equation}
\sigma\left(  u\right)  \equiv\left.  u\left(  t\right)  \right\vert
_{t=\Delta}=\tan\left(  \Delta+\arctan u\right)  =\frac{u+\tan\Delta}%
{1-u\tan\Delta}\ .
\end{equation}
Choosing $\Delta=\pi/4$ for convenience, a quick check on (\ref{BetaFcnlEqn})
gives
\begin{equation}
\beta\left(  u\right)  =1+u^{2}\ ,\ \ \ \sigma\left(  u\right)  =\frac
{1+u}{1-u}\ ,\ \ \ \frac{d\sigma\left(  u\right)  }{du}=\frac{2}{\left(
1-u\right)  ^{2}}\ ,
\end{equation}
So $\beta\left(  \sigma\left(  u\right)  \right)  =\frac{d\sigma\left(
u\right)  }{du}~\beta\left(  u\right)  $ indeed holds. \ In fact, the
functional equation\ obeyed by $\Psi$, namely,
\begin{equation}
e^{\Delta}~\Psi\left(  u\right)  =\Psi\left(  \frac{u+\tan\Delta}%
{1-u\tan\Delta}\right)  \ ,
\end{equation}
actually belongs to the first class of examples discussed in \S 2 of the
original paper by Schr\"{o}der \cite{S}. \ (As an aside, we also note that
(\ref{ModBCSBeta}) is a complexified form of the Beverton--Holt--Skellam model
from population dynamics\ \cite{TLC}.)

The trajectory (\ref{ModBCSTrajectory}) describes a limit cycle because it is
periodic in $t$, as observed by LeClair et al. \cite{LeC} (also see
\cite{GW}). \ This is clearly a consequence of $\Psi^{-1}$ being periodic, as
we remarked earlier in a general context, in \S II. \ However, the
discontinuity in the trajectory is somewhat peculiar to the model. \ This
discontinuity is also displayed by the step scaling function, as shown in the
Figure.%
%TCIMACRO{\FRAME{dtbpFU}{4.9618in}{4.356in}{0pt}{\Qcb{Step-scaling function
%$\sigma\left(  u\right)  =\tan\left(  \frac{\pi}{4}+\arctan u\right)  $, in
%blue, compared to the identity map. \ Note the discontinuity at $u=1$ and the
%lack of a real fixed point.}}{}{bcsdiscontinuity.eps}%
%{\special{ language "Scientific Word";  type "GRAPHIC";
%maintain-aspect-ratio TRUE;  display "USEDEF";  valid_file "F";
%width 4.9618in;  height 4.356in;  depth 0pt;  original-width 5.0274in;
%original-height 4.411in;  cropleft "0";  croptop "1";  cropright "1";
%cropbottom "0";  filename 'BCSDiscontinuity.eps';file-properties "XNPEU";}}}%
%BeginExpansion
\begin{center}
\includegraphics[
height=4.356in,
width=4.9618in
]%
{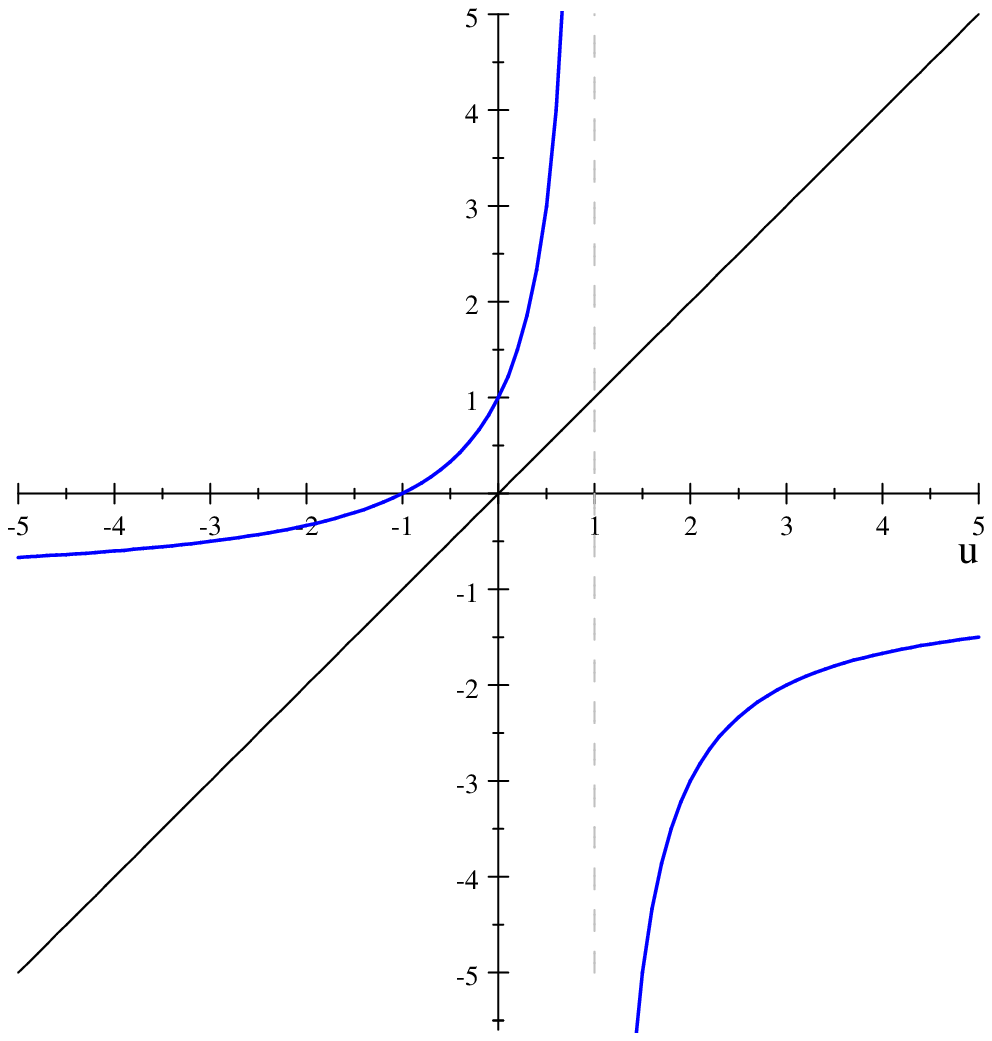}%
\\
Step-scaling function $\sigma\left(  u\right)  =\tan\left(  \frac{\pi}%
{4}+\arctan u\right)  $, in blue, compared to the identity map. \ Note the
discontinuity at $u=1$ and the lack of a real fixed point.
\end{center}
%EndExpansion
Since $\sigma$ does \emph{not} intercept the identity map there is no
\emph{real} fixed point in this case, although $\Psi$ can be constructed by
series solution about the purely imaginary fixed points at $u=\pm i$.

We defer to the detailed discussions in \cite{LeC}\ for a more complete
picture of the physics described by this example. \ We return to a
model-independent viewpoint, to explore other ways that limit cycles might be encountered.

\section{More on the $\beta$ functional equation}

For emphasis, we state again the functional equation relating the
renormalization \textquotedblleft velocity\textquotedblright\ $\beta$ to the
step-scaling function $\sigma$.%
\begin{equation}
\beta\left(  \sigma\left(  u\right)  \right)  =\frac{d\sigma\left(  u\right)
}{du}~\beta\left(  u\right)  \ . \tag{18}\label{BetaFcnlEqnAgain}%
\end{equation}
Generally speaking, if $\beta$ is known, this (nonlinear) functional equation
determines $\sigma\left(  u\right)  $, up to a constant of integration, while
if $\sigma\left(  u\right)  $ is known, the equation determines $\beta$, again
up to a (normalization) constant. \ However, the equation may hold in store
some surprises. \ 

A fixed point $u_{\ast}$ in the step-scaling scheme \emph{must} obey
$\sigma\left(  u_{\ast}\right)  =u_{\ast}$. \ Thus at a fixed point, it is
trivially true that $\beta\left(  \sigma\left(  u_{\ast}\right)  \right)
=\beta\left(  u_{\ast}\right)  $. \ But what has this to do with the standard
fixed point condition that $\beta\left(  u_{\ast}\right)  =0$? \ If $\left.
d\sigma/du\right\vert _{u_{\ast}}\neq1$, then it would \emph{seem} to follow
from the functional equation that $\beta\left(  u_{\ast}\right)  =0 $ when
$\sigma\left(  u_{\ast}\right)  =u_{\ast}$, while if $\left.  d\sigma
/du\right\vert _{u_{\ast}}=1$, the functional equation itself would not lead
to any conclusion about the value of $\beta\left(  u_{\ast}\right)  $. \ But
there is a subtle assumption here: \ All this is true if there is only
\emph{one} \emph{branch} of the analytic function giving rise to $\beta$.
\ Related to this, a \emph{zero} in $d\sigma/du$ may induce, coincide with, or
even supplant a zero of $\beta$. \ This is at odds with the \emph{usual}
renormalization group point of view, but in general this too can occur, and
when it does, it may foreshadow a much richer renormalization structure. \ 

If we have a zero in $d\sigma/du$, say at $u_{0}$, and $\beta\left(
u_{0}\right)  $ is finite, then the functional equation implies $\beta\left(
\sigma\left(  u_{0}\right)  \right)  =0$. \ Now, in the \emph{usual}
renormalization group situation, which we may describe as a purely
\emph{first-order }framework, $u\left(  t\right)  $ is completely determined
by $u\left(  t_{0}\right)  $ and a single function $\beta\left(  u\right)
=du/dt$. \ So, if $\beta\left(  \sigma\left(  u_{0}\right)  \right)  =0$ and
$\sigma\left(  u\right)  $ is just the initial $u$ after having been
$t$-evolved for some discrete step in $t$, this means that $\sigma\left(
u_{0}\right)  =u_{0}$, and therefore $\beta\left(  u_{0}\right)  =0$ must also
hold. \ The fact that $\left.  d\sigma/du\right\vert _{u_{0}}$vanishes does
not make much difference in this purely first-order point of view
\cite{trembling}.

However, in the \emph{quasi-Hamiltonian} framework defined and discussed in
\cite{CZ2}, it may be that $\sigma\left(  u_{0}\right)  $ is not just a zero
of $\beta$, but rather it is a \emph{branch point} of an analytic function
whose various branches constitute a \emph{family} of $\beta$s. \ In this
situation, to completely determine the $t$-evolution of $u$ it is necessary to
specify a transition function. \ That is to say, it is necessary to give a
prescription describing how the trajectory switches from one branch of $\beta$
to another when the branch point $\sigma\left(  u_{0}\right)  $ is
encountered. \ 

In this approach, $\sigma\left(  u_{0}\right)  $ is a \textquotedblleft
turning point\textquotedblright\ in the evolution, and not necessarily a fixed
point. \ Hence it is not necessary for $\sigma\left(  u_{0}\right)  $ to be
the same as $u_{0}$. \ Nor is it necessary for $\beta\left(  u_{0}\right)  $
to vanish.

The quasi-Hamiltonian approach brings to mind a Hamiltonian system whose
underlying dynamics is actually a second-order differential equation, but
which has been reduced to a first-order system between turning points through
the use of energy conservation (see the first Appendix). \ Thus one would
expect, at the very least, that the corresponding $\beta$ function could flip
sign at the turning point $\sigma\left(  u_{0}\right)  $. \ 

In fact, the situation can be more complicated. \ There may be \emph{more}
than the two branch choices $\pm\left\vert \beta\right\vert $ at the turning
point. \ In general there may be an \emph{infinite} number of branches from
which to choose $\beta$! \ 

A local, differential way to think about the various alternatives is to
consider the RG \emph{acceleration}, \emph{jerk}, etc., along the trajectory
as computed through use of the chain rule: \
\begin{equation}
\frac{d^{n}\beta\left(  u\right)  }{dt^{n}}=\beta\left(  u\right)  ~\frac
{d}{du}\left(  \frac{d^{n-1}\beta\left(  u\right)  }{dt^{n-1}}\right)  \ .
\end{equation}
So, for instance, if
\begin{equation}
\beta\left(  u\right)  \underset{u\rightarrow\sigma\left(  u_{0}\right)
}{\sim}\left(  \sigma\left(  u_{0}\right)  -u\right)  ^{p}\ ,
\label{BetaNearTurningPoint}%
\end{equation}
then
\begin{equation}
\frac{d^{n}\beta\left(  u\right)  }{dt^{n}}\underset{u\rightarrow\sigma\left(
u_{0}\right)  }{\sim}\left(  \sigma\left(  u_{0}\right)  -u\right)  ^{\left(
n+1\right)  p-n}\ . \label{BetaDots}%
\end{equation}
Thus, even though $\beta\left(  \sigma\left(  u_{0}\right)  \right)  =0$, the
RG acceleration $d\beta/dt$ can be nonzero at $u=\sigma\left(  u_{0}\right)  $
if $0<p\leq1/2$, indicating that $\sigma\left(  u_{0}\right)  $ is a turning
point and not a fixed point for the continuous flow under these circumstances.

All these things are best understood through consideration of some toy
models.\newpage

\section{Toy models and novel behavior}

Consider a toy example\ that is unphysical (so far as we are aware) but can be
solved in closed form \cite{S}, \
\begin{equation}
\sigma\left(  u\right)  =2u\left(  1-u\right)  \ .
\end{equation}%
%TCIMACRO{\FRAME{dtbpFU}{4.5405in}{3.0246in}{0pt}{\Qcb{Toy model $\sigma\left(
%u\right)  =2u\left(  1-u\right)  $ (solid red) and $\sigma^{-1}\left(
%u\right)  $ (dashed red) versus $u$.}}{}{schroederrenormprd__9.eps}%
%{\special{ language "Scientific Word";  type "GRAPHIC";
%maintain-aspect-ratio TRUE;  display "USEDEF";  valid_file "F";
%width 4.5405in;  height 3.0246in;  depth 0pt;  original-width 4.5981in;
%original-height 3.0539in;  cropleft "0";  croptop "1";  cropright "1";
%cropbottom "0";
%filename 'SchroederRenormPRD__9.eps';file-properties "XNPEU";}}}%
%BeginExpansion
\begin{center}
\includegraphics[
height=3.0246in,
width=4.5405in
]%
{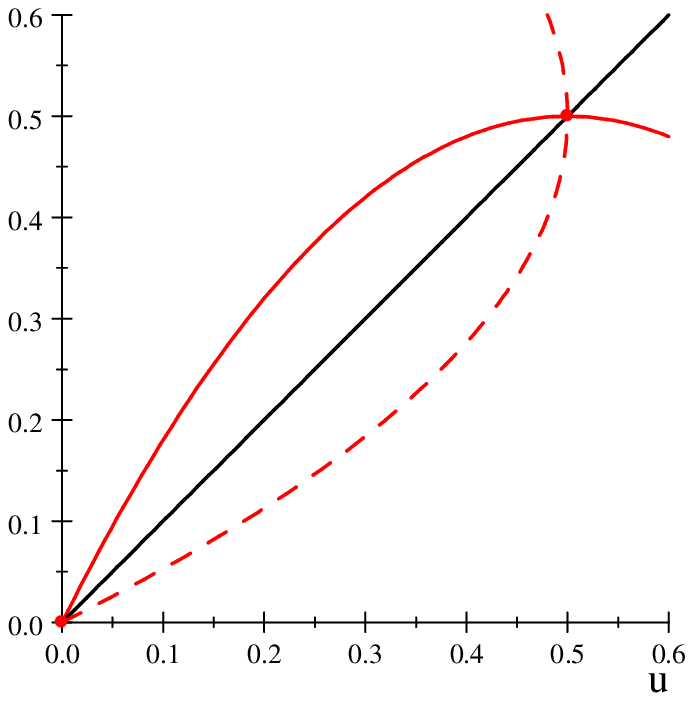}%
\\
Toy model $\sigma\left(  u\right)  =2u\left(  1-u\right)  $ (solid red) and
$\sigma^{-1}\left(  u\right)  $ (dashed red) versus $u$.
\end{center}
%EndExpansion
There are fixed points of $\sigma\left(  u\right)  $\ at $0$ and $u_{\ast
}=1/2$, as indicated by red dots in the Figure. \ The functional equation for
the $\beta$ function in this case is
\begin{equation}
\beta\left(  2u\left(  1-u\right)  \right)  =2\left(  1-2u\right)
\beta\left(  u\right)  \ ,
\end{equation}
with the normalization determined to be $\ln2$ from the eigenvalue of
Schr\"{o}der's equation at $u=0$, namely, $2\Psi\left(  u\right)  =\Psi\left(
2u\left(  1-u\right)  \right)  $. \ Series solution of the functional equation
for $\beta$ about $u=0$ immediately gives%
\begin{equation}
\beta\left(  u\right)  /\ln2=u-u^{2}-\frac{2}{3}u^{3}-\frac{2}{3}u^{4}%
-\frac{4}{5}u^{5}-\frac{16}{15}u^{6}-\frac{32}{21}u^{7}+\cdots=\frac{1}%
{2}\left(  2u-1\right)  \ln\left(  1-2u\right)  \ ,
\end{equation}
where the latter closed form is not difficult to guess from the first twenty
or so terms in the explicit series, and is easily checked to solve the
functional equation exactly. \ This $\beta$ has zeroes precisely at the fixed
points of $\sigma$, again as indicated by red dots in the following Figure.%
%TCIMACRO{\FRAME{dtbpFU}{4.5405in}{3.0246in}{0pt}{\Qcb{The toy $\beta$
%function, $\beta\left(  u\right)  =\frac{1}{2}\left(  \ln2\right)  \left(
%2u-1\right)  \ln\left(  1-2u\right)  $.}}{}{schroederrenormprd__10.eps}%
%{\special{ language "Scientific Word";  type "GRAPHIC";
%maintain-aspect-ratio TRUE;  display "USEDEF";  valid_file "F";
%width 4.5405in;  height 3.0246in;  depth 0pt;  original-width 4.5981in;
%original-height 3.0539in;  cropleft "0";  croptop "1";  cropright "1";
%cropbottom "0";
%filename 'SchroederRenormPRD__10.eps';file-properties "XNPEU";}}}%
%BeginExpansion
\begin{center}
\includegraphics[
height=3.0246in,
width=4.5405in
]%
{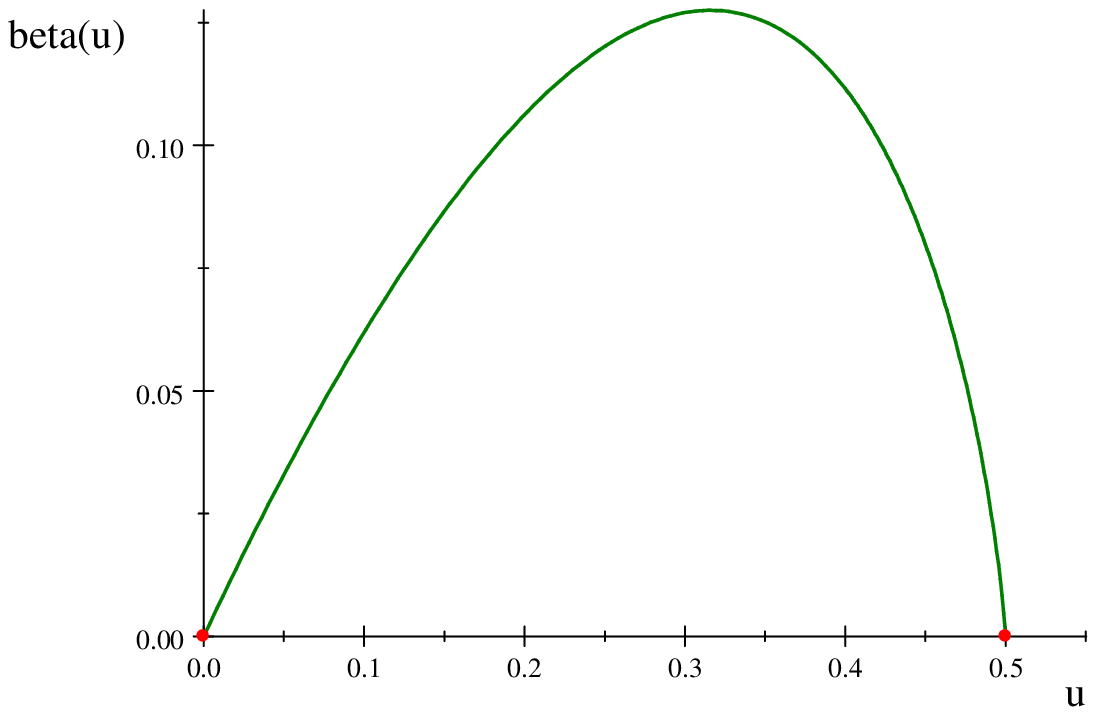}%
\\
The toy $\beta$ function, $\beta\left(  u\right)  =\frac{1}{2}\left(
\ln2\right)  \left(  2u-1\right)  \ln\left(  1-2u\right)  $.
\end{center}
%EndExpansion

For this special case, $d\sigma/du=2\left(  1-2u\right)  $ vanishes at
$u_{\ast}=1/2$, the nontrivial fixed point of $\sigma$, and in addition
$\beta\left(  1/2\right)  =0$, but these coincidences are \emph{not} true in
general, as we shall see. \ In fact, in spite of the divergence in $\left.
d\beta\left(  u\right)  /du\right\vert _{u=u_{\ast}}$, both the RG
acceleration and the RG jerk vanish at the nontrivial fixed point, $\frac
{d}{dt}\beta\left(  u_{\ast}\right)  =0=\frac{d^{2}}{dt^{2}}\beta\left(
u_{\ast}\right)  $, as do all higher $t$ derivatives of $\beta$, since
$\lim_{u\rightarrow1/2}\left(  2u-1\right)  \ln^{n}\left(  1-2u\right)  =0$
for any $n$. \ Thus in this case the RG flow into the nontrivial fixed point
takes place with a very \textquotedblleft soft landing.\textquotedblright

Integration of $du/dt=\beta\left(  u\right)  $ gives%
\begin{equation}
u\left(  t\right)  =\frac{1}{2}\left(  1-\left(  1-2u\left(  0\right)
\right)  ^{2^{t}}\right)  \text{ .}%
\end{equation}
This is precisely of the FC form sanctioned by Schr\"{o}der, namely, $u\left(
t\right)  =\Psi^{-1}\left(  2^{t}\Psi\left(  u\left(  0\right)  \right)
\right)  $, with $\Psi\left(  x\right)  =\ln\left(  1-2x\right)  $ and
$\Psi^{-1}\left(  x\right)  =\frac{1}{2}\left(  1-e^{x}\right)  $. \ A
representative trajectory is shown here.%
%TCIMACRO{\FRAME{dtbpFU}{4.5405in}{3.0246in}{0pt}{\Qcb{A trajectory for the
%$\sigma\left(  u\right)  =2u\left(  1-u\right)  $ model, with $u\left(
%0\right)  =1/4$.}}{}{schroederrenormprd__11.eps}%
%{\special{ language "Scientific Word";  type "GRAPHIC";
%maintain-aspect-ratio TRUE;  display "USEDEF";  valid_file "F";
%width 4.5405in;  height 3.0246in;  depth 0pt;  original-width 4.5981in;
%original-height 3.0539in;  cropleft "0";  croptop "1";  cropright "1";
%cropbottom "0";
%filename 'SchroederRenormPRD__11.eps';file-properties "XNPEU";}}}%
%BeginExpansion
\begin{center}
\includegraphics[
height=3.0246in,
width=4.5405in
]%
{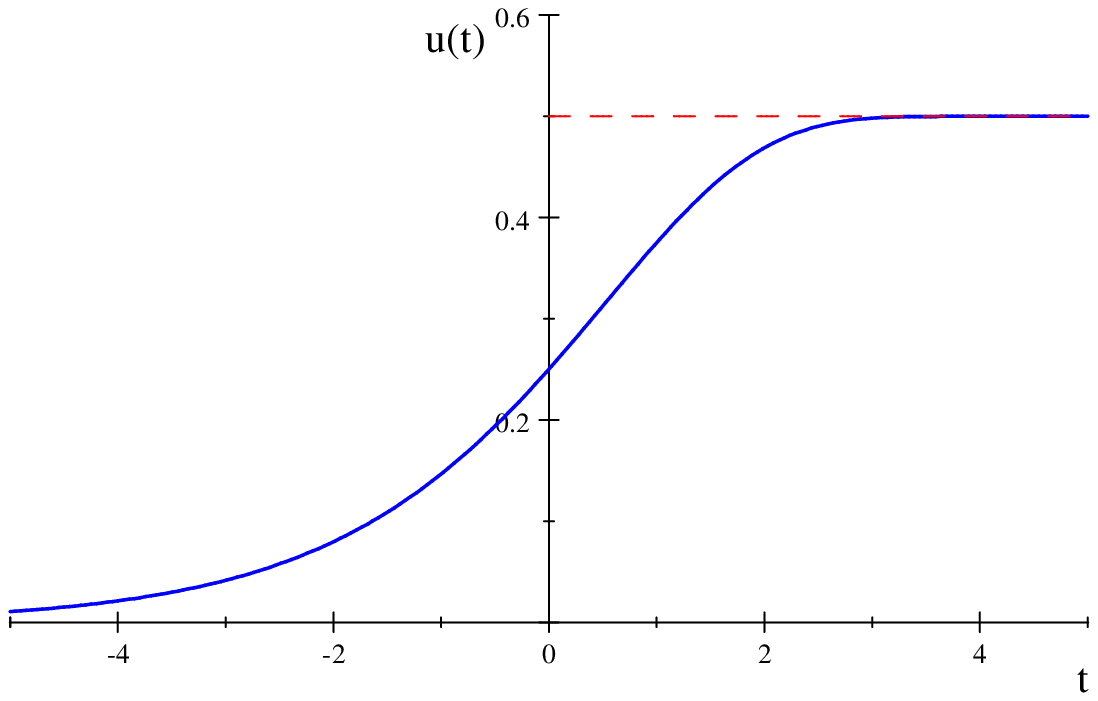}%
\\
A trajectory for the $\sigma\left(  u\right)  =2u\left(  1-u\right)  $ model,
with $u\left(  0\right)  =1/4$.
\end{center}
%EndExpansion

More generally, consider toy models for step-scaling functions based on the
logistic map \cite{CE,D}. \ For parameter $s$, with $0\leq s\leq4$, let
\begin{equation}
\sigma\left(  u,s\right)  =s~u\left(  1-u\right)  \ .
\end{equation}
This $\sigma$ has fixed points at $0$ and $u_{\ast}=1-\frac{1}{s}$. \ Except
for the special case $s=2$, $u_{\ast}$ does \emph{not} coincide with the
location of the maximum of the map where $d\sigma/du=0$, as given by
$\sigma\left(  1/2\right)  =s/4$. \ 

The functional equation for the $\beta$ function in this case is%
\begin{equation}
\beta\left(  su\left(  1-u\right)  ,s\right)  =s\left(  1-2u\right)
\beta\left(  u,s\right)  \ ,
\end{equation}
where the normalization of the solution is given by $\ln\left(  s\right)  $.
\ That is to say, $\beta=\left(  \ln s\right)  ~\Psi/\left(  d\Psi/du\right)
$ where $\Psi$ is the corresponding solution of Schr\"{o}der's equation with
eigenvalue $s$.%
\begin{equation}
s~\Psi\left(  u,s\right)  =\Psi\left(  s~u\left(  1-u\right)  ,s\right)  \ .
\end{equation}
While these functional equations do not admit closed-form solutions, for
generic $s$, they can be solved numerically with sufficient accuracy for our
purposes here through a combination of series and functional methods. \ 

One striking result of this numerical analysis is that $\beta\left(  u_{\ast
}\right)  \neq0$, for $2<s\leq4$, so the fixed point of the step-scaling
function is actually \emph{not} a true fixed point under continuous changes in scale!

Explicit series solution of the functional equation for $\beta$ about $u=0$
gives%
\begin{gather}
\beta\left(  u,s\right)  /\left(  \ln s\right)  =u-\frac{1}{s-1}~u^{2}%
-\frac{2}{s^{2}-1}~u^{3}-\frac{4+5s}{\left(  s^{2}-1\right)  \left(
s^{2}+s+1\right)  }~u^{4}\nonumber\\
-2\frac{4+5s+7s^{2}}{\left(  s^{2}-1\right)  \left(  s^{2}+s+1\right)  \left(
s^{2}+1\right)  }~u^{5}-2\frac{8+18s+31s^{2}+42s^{3}+35s^{4}+21s^{5}}{\left(
s^{2}-1\right)  \left(  s^{2}+1\right)  \left(  s^{4}+s^{3}+s^{2}+s+1\right)
\left(  s^{2}+s+1\right)  }~u^{6}\nonumber\\
-4\frac{8+10s+21s^{2}+25s^{3}+39s^{4}+21s^{5}+33s^{6}}{\left(  s^{2}-1\right)
\left(  s^{2}+1\right)  \left(  s^{4}+s^{3}+s^{2}+s+1\right)  \left(
s^{2}+s+1\right)  \left(  s^{2}-s+1\right)  }~u^{7}+O\left(  u^{8}\right)  \ .
\end{gather}
More generally we write%
\begin{equation}
\beta\left(  u,s\right)  /\left(  \ln s\right)  =u\left(  1+\sum_{n=1}%
^{\infty}c_{n}\left(  s\right)  ~u^{n}\right)  \ ,\ \ \ c_{1}=\frac{1}%
{1-s}\ ,\ \ \ c_{2}=\frac{2}{1-s^{2}}\ , \label{LogisticBetaSeries}%
\end{equation}
with higher coefficients in the series given by the recursion relation (here
$\left\lfloor \cdots\right\rfloor $ is the integer-valued floor function)%
\begin{equation}
c_{n+2}\left(  s\right)  =\frac{1}{1-s^{n+2}}\left(  2c_{n+1}\left(  s\right)
+\sum_{j=\left\lfloor \frac{n+1}{2}\right\rfloor }^{n+1}\binom{j+1}%
{2j-1-n}\left(  -1\right)  ^{n-j}s^{j}c_{j}\left(  s\right)  \right)  \text{
\ \ for }n\geq1\text{.}%
\end{equation}
Numerical study \cite{CV} of $c_{n}\left(  s\right)  $, for various values of
$s$ and $n\leq200$, provides compelling evidence that
(\ref{LogisticBetaSeries}) converges for $\left\vert u\right\vert <R\left(
s\right)  $ where
\begin{equation}
R\left(  s\right)  =\frac{1}{\lim\limits_{n\rightarrow\infty}\sup\left(
\left\vert c_{n}\left(  s\right)  \right\vert ^{1/n}\right)  }=\left\{
\begin{array}
[c]{ccc}%
\dfrac{1}{2} & \text{if} & 0<s\leq\frac{2}{3}\ ,\\
&  & \\
\left\vert 1-\dfrac{1}{s}\right\vert  & \text{if} & \frac{2}{3}\leq
s\leq2\ ,\\
&  & \\
\dfrac{s}{4} & \text{if} & 2\leq s\leq4\ .
\end{array}
\right.  \label{Radii}%
\end{equation}
A closed form is not known for (\ref{LogisticBetaSeries}), except for $s=0,$
$\pm2,$ and $4$. \ Nonetheless, as already mentioned, the model is amenable to
numerical analysis for generic $s$. \ 

In particular, the functional equation can be exploited to continue the series
and exhibit the various branches, $\beta_{n}$, of the multi-valued $\beta$
function that is encountered for $2<s\leq4$ \cite{CV}. \ For example,
\begin{subequations}
\begin{align}
\beta_{0}\left(  u,s\right)   &  =\sqrt{s^{2}-4su}~\beta\left(  \frac{1}%
{2s}\left(  s-\sqrt{s^{2}-4su}\right)  ,s\right)  \ ,\label{PrimaryBranch}\\
\beta_{1}\left(  u,s\right)   &  =-\sqrt{s^{2}-4su}~\beta_{0}\left(  \frac
{1}{2s}\left(  s+\sqrt{s^{2}-4su}\right)  ,s\right)  \ ,
\label{SecondaryBranch}%
\end{align}
where $\beta$ is the explicit series (\ref{LogisticBetaSeries}), $\beta_{0} $
is the continuation of this series through use of the functional equation to
give the principal branch on the interval, $0\leq u\leq s/4$, and $\beta_{1}$
is the first alternative branch which is real-valued on the sub-interval
$\frac{1}{4}s^{2}\left(  1-\frac{1}{4}s\right)  \leq u\leq\frac{s}{4}$. \ Etc.
\ For $s\leq2$ only one branch is needed, namely, $\beta_{0}$, but additional
branches, such as $\beta_{1}$, are required to develop completely the
trajectories for $s>2$. \ In the latter situation, an infinite sequence of
real-valued branch functions is given by iterating the definition of
$\beta_{1}$. \ Thus,
\end{subequations}
\begin{equation}
\beta_{n+1}\left(  u,s\right)  =-\sqrt{s^{2}-4su}~\beta_{n}\left(  \frac
{1}{2s}\left(  s+\sqrt{s^{2}-4su}\right)  ,s\right)  \ .
\label{(n+1)aryBranch}%
\end{equation}
These are all the additional branches of $\beta$ that are needed for
$2<s\leq3$ and for $s=4$, but for $3<s<4$ there are other branch function
sequences that must be taken into account to describe fully the continuous
trajectory $u\left(  t\right)  $. \ 

By construction, stemming from the fact that $\sigma$ is quadratic, for $s>2$
the branches of $\beta$ given by (\ref{(n+1)aryBranch}) have \emph{square-root
zeroes} at points obtained by iterating the action of the step-scaling
function, starting from $u_{0}=1/2$. \ That is, zeroes are given by the
sequence $\left\{  \sigma\left(  u_{0}\right)  ,\sigma\left(  \sigma\left(
u_{0}\right)  \right)  ,\cdots\right\}  $. \ Thus the $\beta_{n}$ branches
arrive at those zeroes with infinite slope, exactly as described by
(\ref{BetaNearTurningPoint}), and its $u$ derivative, with $p=1/2$. \ On the
other hand, the product $\beta~d\beta/du=d\beta/dt$ is finite and nonvanishing
in the limit as $u$ goes to one of these zeroes. \ Consequently, the RG
acceleration does not vanish for points in the sequence of $\beta$ zeroes:
\ They are turning points, not fixed points.

Further general discussion of this class of toy models would take us too far
afield. \ Suffice it to consider here three other explicit examples, two based
on numerical analysis ($s=11/4$ and $s=10/3$), and one based on elementary
closed-form expressions ($s=4$). \ For convenience, we first take\ $s=11/4$.
\ This gives a step-scaling model $\sigma\left(  u\right)  =\frac{11}%
{4}~u\left(  1-u\right)  $, with fixed points at $0$ and $u_{\ast}=7/11$.

Constructing graphs like those for the previous toy model, we find that there
are several interesting differences between the Figures for the two models.
\ The slope of the step-scaling function is now negative at the nontrivial
fixed point, as opposed to the vanishing slope of the previous toy model.
\ This negative slope is a crucial ingredient that gives rise to the
multi-valued-ness of the $\beta$ function, as shown in the Figures.%
%TCIMACRO{\FRAME{dtbpFU}{4.5405in}{3.0246in}{0pt}{\Qcb{Toy model $\sigma\left(
%u\right)  =\frac{11}{4}~u\left(  1-u\right)  $ (solid red) and $\sigma
%^{-1}\left(  u\right)  $ (dashed red) versus $u$.}}{}%
%{schroederrenormprd__12.eps}{\special{ language "Scientific Word";
%type "GRAPHIC";  maintain-aspect-ratio TRUE;  display "USEDEF";
%valid_file "F";  width 4.5405in;  height 3.0246in;  depth 0pt;
%original-width 4.5981in;  original-height 3.0539in;  cropleft "0";
%croptop "1";  cropright "1";  cropbottom "0";
%filename 'SchroederRenormPRD__12.eps';file-properties "XNPEU";}}}%
%BeginExpansion
\begin{center}
\includegraphics[
height=3.0246in,
width=4.5405in
]%
{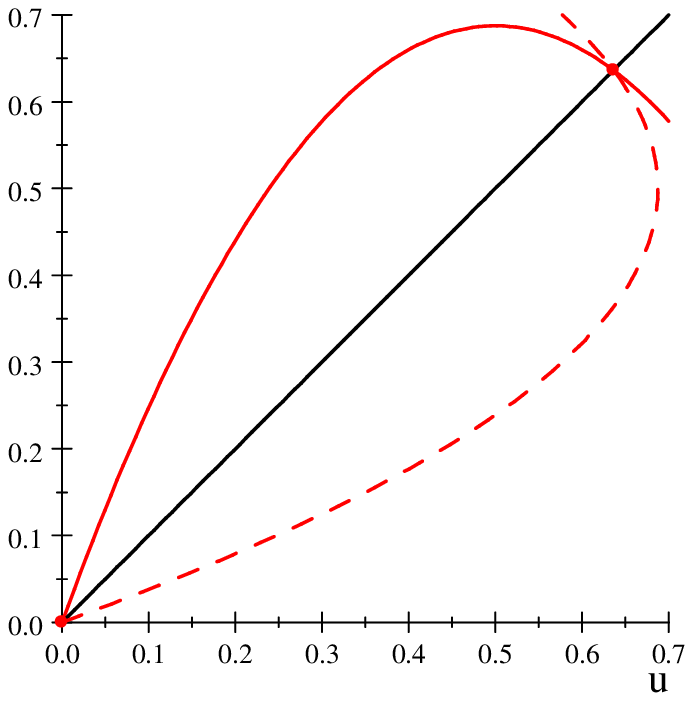}%
\\
Toy model $\sigma\left(  u\right)  =\frac{11}{4}~u\left(  1-u\right)  $ (solid
red) and $\sigma^{-1}\left(  u\right)  $ (dashed red) versus $u$.
\end{center}
%EndExpansion%
%TCIMACRO{\FRAME{dtbpFU}{4.5405in}{3.0246in}{0pt}{\Qcb{Six branches of the
%multi-valued $\beta$ function for the model with $\sigma\left(  u\right)
%=\frac{11}{4}~u\left(  1-u\right)  $.}}{}{schroederbeta11fourths.eps}%
%{\special{ language "Scientific Word";  type "GRAPHIC";
%maintain-aspect-ratio TRUE;  display "USEDEF";  valid_file "F";
%width 4.5405in;  height 3.0246in;  depth 0pt;  original-width 4.5981in;
%original-height 3.0539in;  cropleft "0";  croptop "1";  cropright "1";
%cropbottom "0";
%filename 'SchroederBeta11Fourths.eps';file-properties "XNPEU";}}}%
%BeginExpansion
\begin{center}
\includegraphics[
height=3.0246in,
width=4.5405in
]%
{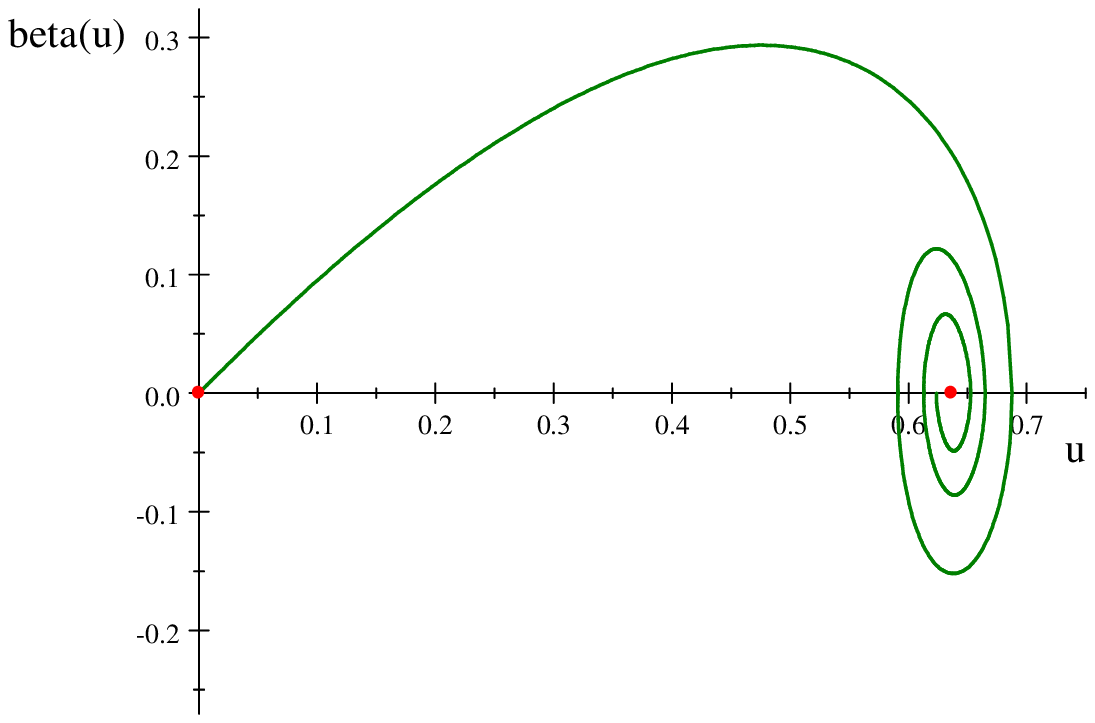}%
\\
Six branches of the multi-valued $\beta$ function for the model with
$\sigma\left(  u\right)  =\frac{11}{4}~u\left(  1-u\right)  $.
\end{center}
%EndExpansion

To avoid confusion, we emphasize that this last \textquotedblleft
phase-space\textquotedblright\ Figure shows the real-valued branches of
$\beta\left(  u\right)  $, i.e. the Figure does \textbf{not} show a trajectory
in a two-dimensional coupling space. \ The trajectory $u\left(  t\right)  $ is
\emph{one}-dimensional, just as it was in the previous toy model, as
illustrated in the following graph. \ Although $u_{\ast}=7/11$ is a fixed
point of $\sigma$, it is \emph{not} a zero of $\beta$ on \emph{any} of its
real-valued branches. \ Moreover, at the nontrivial zeroes of $\beta$ the RG
acceleration does not vanish because the relevant branch of $\beta$ has a
compensating infinite slope, as discussed more generally following
(\ref{(n+1)aryBranch}), and as is evident in the plot of the toy trajectory,
so these zeroes are indeed turning points. \ Nevertheless, the net effect of
the multiple branches of $\beta$ on the trajectory $u\left(  t\right)  $ is to
produce an oscillatory convergence to $u_{\ast}=7/11$ as $t\rightarrow\infty$.
\ This is distinctly different from the monotonic approach to $u_{\ast}=1/2$
of the previous toy model whose $\beta$ function had only one real branch.%
%TCIMACRO{\FRAME{dtbpFU}{4.5405in}{3.0246in}{0pt}{\Qcb{A trajectory with
%$u\left(  0\right)  =1/4$, for the $s=11/4$ multi-valued $\beta$ function.}}%
%{}{schroedertrajectory11fourths.eps}{\special{ language "Scientific Word";
%type "GRAPHIC";  maintain-aspect-ratio TRUE;  display "USEDEF";
%valid_file "F";  width 4.5405in;  height 3.0246in;  depth 0pt;
%original-width 4.5981in;  original-height 3.0539in;  cropleft "0";
%croptop "1";  cropright "1";  cropbottom "0";
%filename 'SchroederTrajectory11Fourths.eps';file-properties "XNPEU";}}}%
%BeginExpansion
\begin{center}
\includegraphics[
height=3.0246in,
width=4.5405in
]%
{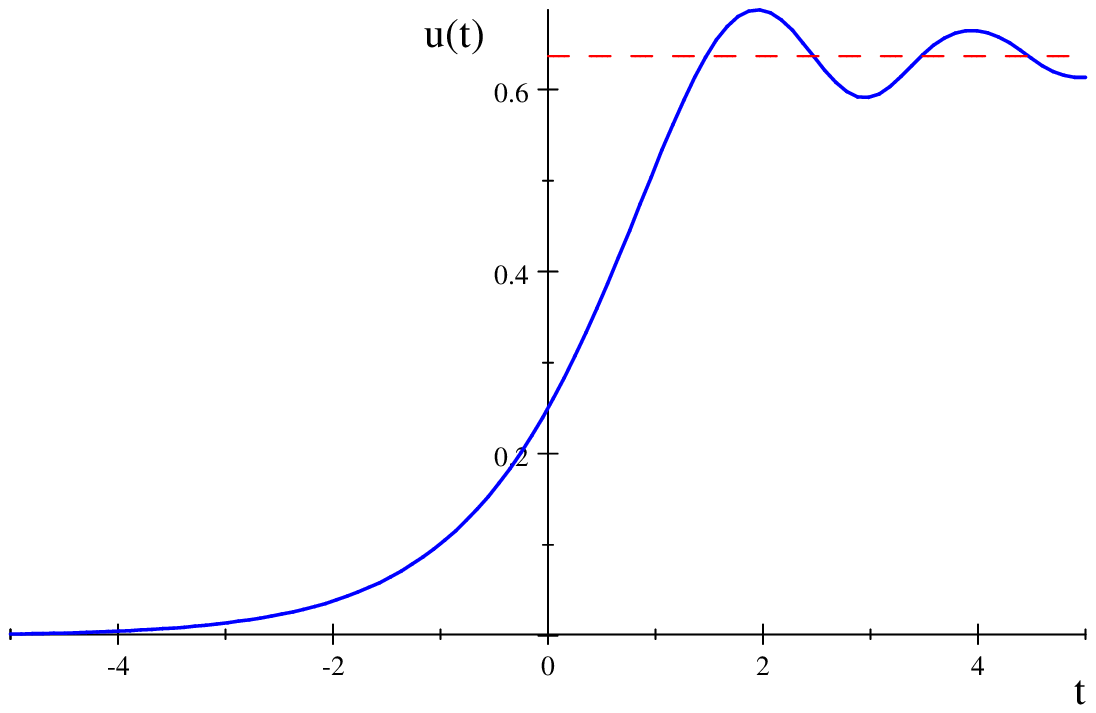}%
\\
A trajectory with $u\left(  0\right)  =1/4$, for the $s=11/4$ multi-valued
$\beta$ function.
\end{center}
%EndExpansion

Next, consider $s=10/3$, another case which yields to numerical analysis but
cannot be described with closed-form results. \ The step-scaling function in
this case is $\sigma\left(  u\right)  =\frac{10}{3}~u\left(  1-u\right)  $,
with fixed points at $0$ and $u_{\ast}=7/10$. \ The graph of $\sigma$ is
similar to the previous $s=11/4$ example, and is left to the reader to plot.
\ This toy example is especially interesting in that it provides an explicit
RG realization of a \emph{limit cycle}, a possibility conjectured by Wilson in
his early, classic study \cite{W}, but not thought to be possible for a model
with only one coupling until relatively recently \cite{WG}. \ Here, the exact
cycle is only realized \emph{asymptotically} in the limit of very large $t$. \ 

The basic mechanism whereby this is achieved for the $s=10/3$ case is the same
as in the previous toy model: \ There are an infinite number of branches of
the underlying analytic $\beta$ function, and the evolving RG trajectory
switches from one branch to another when turning points are encountered. \ The
branch structure is more complicated for $s=10/3$ than for $s=11/4$, however,
with the branch points accumulating around $u_{\text{low}}=\frac{13}{20}%
-\frac{1}{20}\sqrt{13}=0.4697$ and $u_{\text{high}}=\frac{1}{20}\sqrt
{13}+\frac{13}{20}=0.8303$ in such a way that, as $t\rightarrow\infty$, the
trajectory approaches a rectangular sequence of steps between these two
values, hence giving rise to a two-cycle asymptotically in $t$. \ The initial
stages of this large $t$ behavior are evident in the following sample
trajectory.%
%TCIMACRO{\FRAME{dtbpFU}{5.6722in}{3.0273in}{0pt}{\Qcb{A trajectory with
%$u\left(  0\right)  =1/4$, for the $s=10/3$ multi-valued $\beta$ function with
%a 2-cycle limit.}}{}{inverseschroederplot10thirds__1.eps}%
%{\special{ language "Scientific Word";  type "GRAPHIC";
%maintain-aspect-ratio TRUE;  display "USEDEF";  valid_file "F";
%width 5.6722in;  height 3.0273in;  depth 0pt;  original-width 5.7991in;
%original-height 3.0823in;  cropleft "0";  croptop "0.9914";
%cropright "0.9918";  cropbottom "0";
%filename 'InverseSchroederPlot10Thirds__1.eps';file-properties "XNPEU";}}}%
%BeginExpansion
\begin{center}
\includegraphics[
trim=0.000000in 0.000000in 0.047553in 0.026508in,
height=3.0273in,
width=5.6722in
]%
{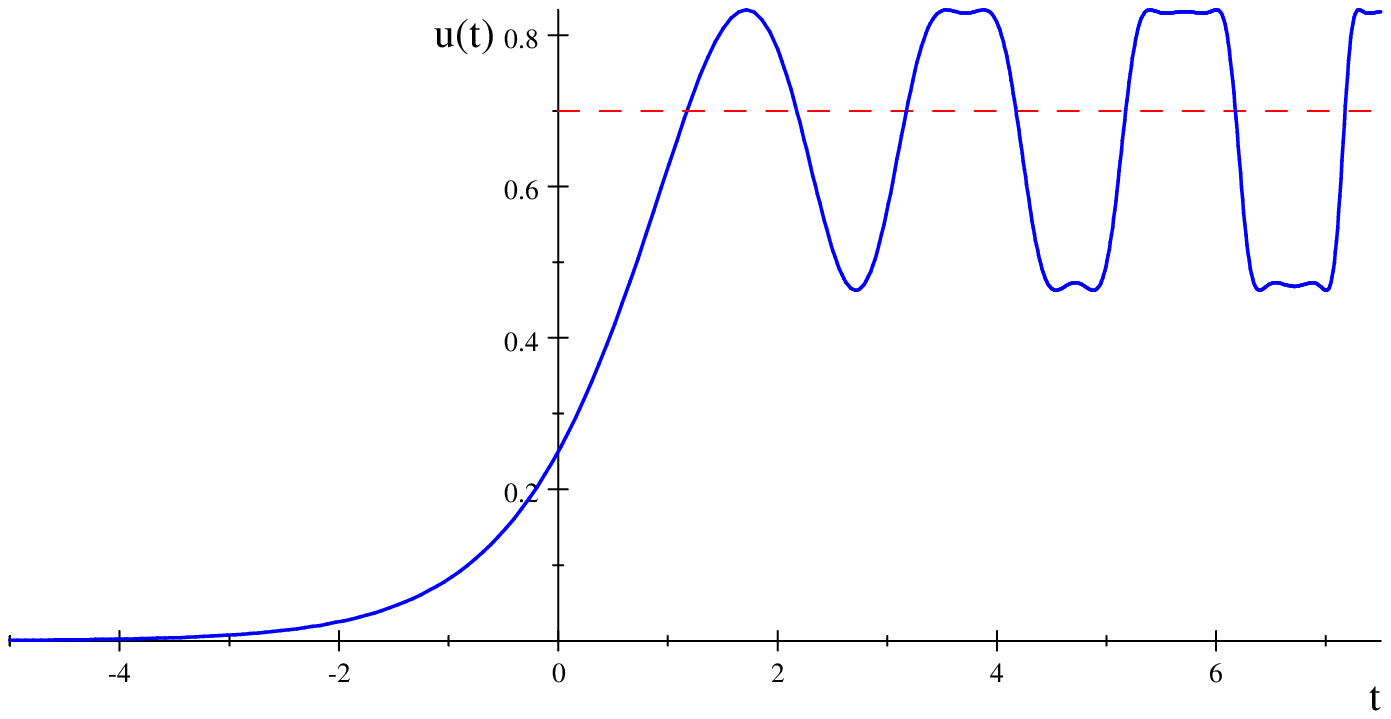}%
\\
A trajectory with $u\left(  0\right)  =1/4$, for the $s=10/3$ multi-valued
$\beta$ function with a 2-cycle limit.
\end{center}
%EndExpansion

While the methods used in the previous example are also applicable to this
case (see (\ref{PrimaryBranch}), (\ref{SecondaryBranch}),
(\ref{(n+1)aryBranch}), and generalizations), we will not construct here the
actual branches of $\beta$ for $s=10/3$, nor will we enumerate the sequencing
of the various branches as the trajectory evolves. \ This information can be
found,\ in a different context, in \cite{CV}. \ Suffice it to say that
although $u_{\ast}=7/10$ is a fixed point of $\sigma$, once again it is
\emph{not} a zero of $\beta$ on \emph{any} of its real-valued branches.

For a final, \emph{chaotic} example, which nevertheless admits an exact,
closed-form solution \cite{S}, consider%
\begin{equation}
\sigma\left(  u\right)  =4u\left(  1-u\right)  \ ,
\end{equation}
for $0\leq u\leq1$. \ In this case the real-valued branches of the analytic
$\beta$ function are given by%
\begin{equation}
\beta_{n}\left(  u\right)  =\left(  \ln4\right)  \sqrt{u\left(  1-u\right)
}\left(  \left(  -1\right)  ^{n}\left\lfloor \frac{1+n}{2}\right\rfloor
\pi+\arcsin\sqrt{u}\right)  \ ,
\end{equation}
for all integer $n\geq0$. \ The functional equation is indeed obeyed, in the
following sense:
\begin{subequations}
\begin{align}
\beta_{n}\left(  \sigma\left(  u\right)  \right)   &  =\beta_{n}\left(
u\right)  ~d\sigma\left(  u\right)  /du\text{ \ \ for \ \ }0\leq u\leq1/2\ ,\\
\beta_{n+1}\left(  \sigma\left(  u\right)  \right)   &  =\beta_{n}\left(
u\right)  ~d\sigma\left(  u\right)  /du\text{ \ \ for \ \ }1/2\leq u\leq1\ .
\end{align}
Note that $\sigma\left(  u\right)  $ approaches the branch point at $\sigma=1$
as $u$ goes to $1/2$. \ Beyond this, for $1/2\leq u\leq1$, the $\beta\left(
\sigma\left(  u\right)  \right)  $ term in (\ref{BetaFcnlEqn}) switches from
the $n$th branch to the $\left(  n+1\right)  $st. \ Correspondingly, at the
nontrivial fixed point $u_{\ast}=3/4$ of the step-scaling function, we see
that $\beta\left(  u_{\ast}\right)  $ does \emph{not} vanish for \emph{any} of
the branches of $\beta$. \ A typical RG trajectory in this chaotic case is
shown here.%
%TCIMACRO{\FRAME{dtbpFU}{4.5405in}{3.0246in}{0pt}{\Qcb{A trajectory with
%$u\left(  0\right)  =1/4$, for the $s=4$ multi-valued $\beta$ function leading
%to chaotic evolution.}}{}{schroederrenormprd__13.eps}%
%{\special{ language "Scientific Word";  type "GRAPHIC";
%maintain-aspect-ratio TRUE;  display "USEDEF";  valid_file "F";
%width 4.5405in;  height 3.0246in;  depth 0pt;  original-width 4.5981in;
%original-height 3.0539in;  cropleft "0";  croptop "1";  cropright "1";
%cropbottom "0";
%filename 'SchroederRenormPRD__13.eps';file-properties "XNPEU";}}}%
%BeginExpansion
\begin{center}
\includegraphics[
height=3.0246in,
width=4.5405in
]%
{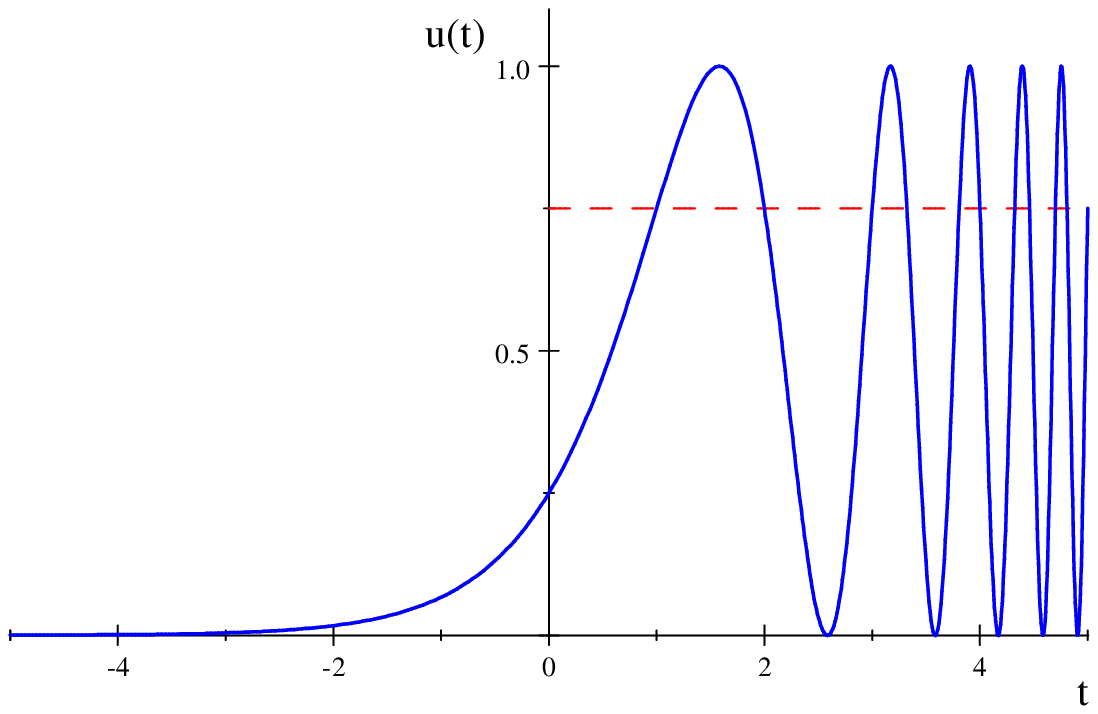}%
\\
A trajectory with $u\left(  0\right)  =1/4$, for the $s=4$ multi-valued
$\beta$ function leading to chaotic evolution.
\end{center}
%EndExpansion
There is one highly unusual feature for the trajectory shown, as well as for
all other trajectories for this example: \ The coupling actually goes to zero
with increasing frequency but always \textquotedblleft bounces
back\textquotedblright\ to positive values. \ For a thorough discussion of
this model using functional methods, in the context of classical mechanics for
a chaotic dynamical system, see \cite{CZ2}.

\section{Discussion}

In this paper we have elucidated, explored, and applied the underlying
functional conjugacy structure of the renormalization group implicit in
Gell-Mann and Low's finite renormalization group equation \cite{GML},
structure which is normally overshadowed by local differential (Lie algebraic)
features. \ We introduced methods to extract the links among fixed points of
exemplary RG trajectories in section II. \ We applied some of these functional
methods to obtain continuous flows from step-scaling functions utilized in
lattice gauge theory, in section III. \ We explained in section IV how novel
features could arise in general. \ We illustrated such features --- including
multi-valued $\beta$ functions, limit cycles, and chaotic trajectories ---
using toy models based on the logistic equation in section V.

The functional conjugacy core of the RG, (\ref{global}), amounts to a solution
by the method of characteristics, (\ref{characteristics}) et seq., so that
scale changes are equivalent to variations of the initial coupling data.
\ This controlling characteristic structure implies nonlocal associations in
renormalization flows. \ In the large, the resulting functionally conjugate RG
trajectories possess an elegant global mathematical consistency which sheds
light on the flow of couplings in the presence of both UV and IR fixed points,
and which also shows that trajectories may admit turning points, i.e. zeros of
$\beta$ functions where not all $t$ derivatives of $\beta$ vanish as a
consequence of intricate branch structure. \ This latter possibility may lead
to limit cycle or even chaotic behavior, as we have illustrated in some
detail. \ 

In this paper, models with only a single RG coupling flow have been considered
. \ Extensions of the functional methods to models with more than one flowing
coupling have yet to be carried out in complete detail. \ This is an area that
warrants further exploration.

While we have not yet fully explored realistic quantum field theoretical
models evincing \emph{all} of the possibilities that we have discussed, we are
confident that they do exist, at least for many of the features we have
described. \ Presumably, the more exotic, \emph{oscillatory} RG flows would
only emerge in systems where the c-theorem \cite{Z,Cardy} (or its equivalent
in higher dimensions \cite{Myers}) does not hold \cite{WG,LeCS}. \ We would be
most pleased to find in actual physical systems the full range of behavior
illustrated by the toy models.
\end{subequations}
\begin{acknowledgments}
\textit{We thank Don Sinclair and Seth Quackenbush for comments on this work,
and an anonymous referee for suggestions to improve the presentation.}
\ \textit{One of us (TC) thanks the CERN Theoretical Physics Group for its
gracious hospitality and generous support, and the paragliders on the
Sal\`{e}ve for a convincing demonstration that interesting things can be done
even in a crowded field. \ The numerical calculations and graphics in this
paper were made using Maple}$^{\textregistered}$\textit{, Mathematica}%
$^{\textregistered}$, \textit{and MuPAD}$^{\textregistered}$\textit{. \ This
work was supported in part by NSF Award 0855386, and in part by the U.S.
Department of Energy, Division of High Energy Physics, under contract
DE-AC02-06CH11357.}
\end{acknowledgments}

\section{Appendix: \ Newtonian trajectory as a transport of data}

Here we discuss analogies between RG flow and one-dimensional motion of a
classical particle, thereby making contact with our previous work
\cite{CZ1,CZ2,CV}.

For fixed energy, $E$, the trajectory $x\left(  t,x_{0}\right)  $ of a
particle moving in a one-dimensional potential, between turning points,
depends only on time $t$ and initial position $x_{0}=\left.  x\right\vert
_{t=0}$, since the initial momentum is fixed by $E$ up to a choice of branch
for $\sqrt{E-V\left(  x\right)  }$ (usually just an overall $\pm$ sign if $V$
itself has only one branch). \ Thus
\begin{equation}
\int_{x_{0}}^{x\left(  t,x_{0}\right)  }\frac{dx}{v\left(  x\right)  }=t\ ,
\label{Time}%
\end{equation}
where $v\left(  x\right)  $ is the velocity profile along the trajectory. \ In
this context it is clear that $v=0$ is not necessarily a fixed point. \ Rather
more often it is a \emph{turning point} of the motion.

Now, here's a simple technique that appears in Gell-Mann and Low (see Eq
(B20), \cite{GML}): \ Differentiate (\ref{Time}) with respect to the initial
position, regarding $t$ and $x_{0}$ as independent variables. \ This gives%
\begin{equation}
\frac{1}{v\left(  x\left(  t,x_{0}\right)  \right)  }\frac{\partial x\left(
t,x_{0}\right)  }{\partial x_{0}}-\frac{1}{v\left(  x_{0}\right)  }=0\ .
\end{equation}
On the other hand, $v\left(  x\left(  t,x_{0}\right)  \right)  =\partial
x\left(  t,x_{0}\right)  /\partial t$, so this last result is just%
\begin{equation}
\frac{\partial x\left(  t,x_{0}\right)  }{\partial t}=v\left(  x_{0}\right)
~\frac{\partial x\left(  t,x_{0}\right)  }{\partial x_{0}}\ .
\label{Transport}%
\end{equation}
That is to say, for want of a better name, this is a one-dimensional
\textquotedblleft Gell-Mann--Low transport equation\textquotedblright\ for
$x\left(  t,x_{0}\right)  $. \ It is not quite \textquotedblleft
advection\textquotedblright\ \cite{Co} since that would have the form of a
conservation law, namely, $\frac{\partial}{\partial t}f=\frac{\partial
}{\partial x}\left(  vf\right)  $. \ The two types of transport are exactly
the same only for constant $v$.

In fact, (\ref{Transport}) is simpler than advection, with solutions of the FC
form%
\begin{equation}
x\left(  t,x_{0}\right)  =\Psi^{-1}\left(  e^{t}\Psi\left(  x_{0}\right)
\right)  \ , \label{Solution}%
\end{equation}
for an appropriately defined Schr\"{o}der function $\Psi$\ (although this
terminology is \emph{not} used in \cite{GML}) with inverse function $\Psi
^{-1}$. \ By \textquotedblleft appropriately defined\textquotedblright\ we
mean,
\begin{equation}
v\left(  x_{0}\right)  =\Psi\left(  x_{0}\right)  /\Psi^{\prime}\left(
x_{0}\right)  =\frac{1}{\left(  \partial\ln\Psi\left(  x_{0}\right)  /\partial
x_{0}\right)  }\ , \label{vPsi}%
\end{equation}
hence $\Psi$\ is essentially just the exponentiated time to reach $x$ from
some reference point,%
\begin{equation}
\Psi\left(  x\right)  =\Psi\left(  x_{\text{ref}}\right)  \exp\left(
\int_{x_{\text{ref}}}^{x}\frac{dy}{v\left(  y\right)  }\right)  \ .
\label{SchroederFunction}%
\end{equation}
Another way to express the result (\ref{Solution}) is as a formal Taylor
series in $t$, rewritten in terms of $v\left(  x_{0}\right)  $ and its
derivatives through the use of (\ref{Transport}). \ Thus%
\begin{align}
x\left(  t,x_{0}\right)   &  =\sum_{n=0}^{\infty}\frac{1}{n!}~t^{n}\left.
\frac{\partial^{n}}{\partial\tau^{n}}x\left(  \tau,x_{0}\right)  \right\vert
_{\tau=0}=\sum_{n=0}^{\infty}\frac{1}{n!}~t^{n}\left(  v\left(  x_{0}\right)
~\frac{\partial}{\partial x_{0}}\right)  ^{n}x_{0}\ \nonumber\\
&  =x_{0}+t~v\left(  x_{0}\right)  +\sum_{n=2}^{\infty}\frac{1}{n!}%
~t^{n}\left(  v\left(  x_{0}\right)  ~\frac{\partial}{\partial x_{0}}\right)
^{n-1}v\left(  x_{0}\right)  \ . \label{series}%
\end{align}
(See Eq (B21) in \cite{GML}.) \ If it is not already evident why this solution
is formally the same as (\ref{Solution}), then repeat some steps from the text
and write the series in (\ref{series}) as exponentiated operators.%
\begin{equation}
x\left(  t,x_{0}\right)  =\left.  e^{t\frac{\partial}{\partial\tau}}~x\left(
\tau,x_{0}\right)  \right\vert _{\tau=0}=e^{t~v\left(  x_{0}\right)
~\frac{\partial}{\partial x_{0}}}~x_{0}=e^{t~\frac{\partial}{\partial\ln
\Psi\left(  x_{0}\right)  }}~x_{0}\ , \label{exponentiated}%
\end{equation}
where in the last step we have used (\ref{vPsi}). \ But now, $x_{0}=\Psi
^{-1}\left(  \Psi\left(  x_{0}\right)  \right)  =\Psi^{-1}\left(  \exp\left(
\ln\Psi\left(  x_{0}\right)  \right)  \right)  $, so the last expression in
(\ref{exponentiated}) reduces to a translation of the variable $\ln\left(
\Psi\left(  x_{0}\right)  \right)  $.
\begin{equation}
e^{t~\frac{\partial}{\partial\ln\Psi\left(  x_{0}\right)  }}~\Psi^{-1}\left(
\exp\left(  \ln\Psi\left(  x_{0}\right)  \right)  \right)  =\Psi^{-1}\left(
\exp\left(  t+\ln\Psi\left(  x_{0}\right)  \right)  \right)  =\Psi^{-1}\left(
e^{t}\Psi\left(  x_{0}\right)  \right)  \ .
\end{equation}
Thus we recover (\ref{Solution}) as an element of an abelian Lie group, given
as usual by exponentiating an element of the underlying algebra ---\ a
symplectomorphism in the present context of classical mechanics.

\section{Appendix: A variable change at a lattice fixed point}

For a supposed nontrivial fixed point, $u_{\ast}=1/g_{\ast}^{2}$, determined
by
\begin{equation}
\sigma\left(  u_{\ast}\right)  =u_{\ast}\ ,
\end{equation}
with $\beta\left(  u_{\ast}\right)  =0$, we change variables in the lattice
data formulas (\ref{u}) and (\ref{sigma})\ of the text, by writing
\begin{equation}
\frac{1}{s}=\left(  1+r\right)  g_{0\ast}^{2}\ ,
\end{equation}
so that $r>0$ for bare couplings $g_{0}^{2}$ \emph{above} the fixed point
value $g_{0\ast}^{2}$. \ Under this change of parameterization, we have%
\begin{align}
u  &  =\frac{1}{\left(  1+r\right)  g_{0\ast}^{2}}\left(  1-\sum_{j=1}%
^{n}c_{j}\left(  \ell\right)  \left(  1+r\right)  ^{j}g_{0\ast}^{2j}\right)
=u_{\ast}+\sum_{j\geq1}\mathfrak{a}_{j}\left(  \ell\right)  r^{j}%
\ ,\label{fixed}\\
u_{\ast}  &  =\frac{1}{g_{\ast}^{2}}=\frac{1}{g_{0\ast}^{2}}\left(
1-\sum_{j\geq1}c_{j}\left(  \ell\right)  g_{0\ast}^{2j}\right)  \ .\\
\frac{du\left(  s\right)  }{ds}  &  =\sum_{j\geq0}\mathfrak{a}_{j}^{\prime
}\left(  \ell\right)  r^{j}\ ,\ \ \ \mathfrak{a}_{j}^{\prime}=\left(
j+1\right)  \mathfrak{a}_{j+1}\frac{dr}{ds}\ ,\ \ \ \frac{dr}{ds}=-\frac
{1}{g_{0\ast}^{2}s^{2}}\ .
\end{align}%
\begin{gather}
\mathfrak{a}_{1}=\frac{1}{g_{0\ast}^{2}}\left(  -1-\sum_{j\geq2}\left(
j-1\right)  c_{j}\left(  \ell\right)  g_{0\ast}^{2j}\right)
\ ,\ \ \ \mathfrak{a}_{2}=\frac{1}{g_{0\ast}^{2}}\left(  1-\frac{1}{2}%
\sum_{j\geq3}\left(  j-1\right)  \left(  j-2\right)  c_{j}\left(  \ell\right)
g_{0\ast}^{2j}\right)  \ ,\\
\mathfrak{a}_{k<\max\left(  n\right)  }=\frac{1}{g_{0\ast}^{2}}\left(  \left(
-1\right)  ^{k}-\frac{1}{k!}\sum_{j\geq k+1}\left(  j-1\right)  \left(
j-2\right)  \cdots\left(  j-k\right)  c_{j}\left(  \ell\right)  g_{0\ast}%
^{2j}\right)  \ ,\ \ \ \mathfrak{a}_{k}\underset{k\geq\max\left(  n\right)
}{=}\frac{\left(  -1\right)  ^{k}}{g_{0\ast}^{2}}\text{ ,}%
\end{gather}
and\ therefore $\frac{1}{g_{0\ast}^{2}}\sum_{j=n+1}^{\infty}\left(  -1\right)
^{j}r^{j}=\frac{r^{n+1}}{\left(  1+r\right)  g_{0\ast}^{2}}$, as it should.
\ Similarly%
\begin{equation}
\sigma\left(  u\right)  =u_{\ast}+\sum_{j\geq1}\mathfrak{a}_{j}\left(
L\right)  r^{j}\ ,\ \ \ \frac{d\sigma\left(  u\left(  s\right)  \right)  }%
{ds}=\sum_{j\geq0}\mathfrak{a}_{j}^{\prime}\left(  L\right)  r^{j}\ .
\end{equation}
Discarding an overall $\frac{dr}{ds}$, the functional equation for the $\beta$
function (\ref{BetaFcnlEqnParametric}) is then
\begin{equation}
\left(  \sum_{j\geq0}\left(  j+1\right)  \mathfrak{a}_{j+1}\left(
\ell\right)  r^{j}\right)  ~\beta\left(  u_{\ast}+\sum_{k\geq1}\mathfrak{a}%
_{k}\left(  L\right)  r^{k}\right)  =\left(  \sum_{j\geq0}\left(  j+1\right)
\mathfrak{a}_{j+1}\left(  L\right)  r^{j}\right)  ~\beta\left(  u_{\ast}%
+\sum_{k\geq1}\mathfrak{a}_{k}\left(  \ell\right)  r^{k}\right)  \ .
\label{SolveThisBySeries}%
\end{equation}
Now it is straightforward albeit tedious to construct, up to an overall
normalization, a series solution in $r$ of this functional equation to obtain
an expression for $\beta$ near the fixed point. \ We shall \emph{assume}
$u_{\ast}$ is a first-order zero of $\beta$. \
\begin{equation}
\beta=\sum_{i=1}^{\min\left(  n,N\right)  }\mathfrak{b}_{i}r^{i}\ .
\end{equation}
It would be nothing more than wishful thinking to carry the series
approximation for $\beta$ beyond that for either $u$ or $\sigma\left(
u\right)  $. \ 

For numerical work described in the text, we build the series solution to
fourth order in $r$. $\ $Define $\mathfrak{a}_{j}\left(  \ell\right)
=\mathfrak{a}_{j}$, $\mathfrak{a}_{j}\left(  L\right)  =\mathfrak{A}_{j}$,
and
\begin{equation}
\beta\left(  u_{\ast}+w\right)  =\mathfrak{b}_{1}w+\mathfrak{b}_{2}%
w^{2}+\mathfrak{b}_{3}w^{3}+\mathfrak{b}_{4}w^{4}\ .
\label{BetaNearLatticeFixedPoint}%
\end{equation}
Then (\ref{SolveThisBySeries}) gives
\begin{align}
&  \left(  \mathfrak{a}_{1}+2\mathfrak{a}_{2}r+3\mathfrak{a}_{3}%
r^{2}+4\mathfrak{a}_{4}r^{3}+5\mathfrak{a}_{5}r^{4}\right)  ~\beta\left(
u_{\ast}+\mathfrak{A}_{1}r+\mathfrak{A}_{2}r^{2}+\mathfrak{A}_{3}%
r^{3}+\mathfrak{A}_{4}r^{4}\right) \nonumber\\
&  =\left(  \mathfrak{A}_{1}+2\mathfrak{A}_{2}r+3\mathfrak{A}_{3}%
r^{2}+4\mathfrak{A}_{4}r^{3}+5\mathfrak{A}_{5}r^{4}\right)  ~\beta\left(
u_{\ast}+\mathfrak{a}_{1}r+\mathfrak{a}_{2}r^{2}+\mathfrak{a}_{3}%
r^{3}+\mathfrak{a}_{4}r^{4}\right)  \ .
\end{align}
Now, we obviously cannot determine the overall normalization of $\beta$ from
the functional equation (\ref{SolveThisBySeries}). \ This fact is why we find
the same coefficient of $r^{1}$ on the LHS and RHS of this last equation, for
any $\mathfrak{b}_{1}$. \ But higher powers of $r$ give nontrivial
information. \ For example, $r^{2}$ gives%
\begin{equation}
\frac{\mathfrak{b}_{2}}{\mathfrak{b}_{1}}=\frac{\mathfrak{A}_{1}%
\mathfrak{a}_{2}-\mathfrak{A}_{2}\mathfrak{a}_{1}}{\mathfrak{A}_{1}%
\mathfrak{a}_{1}\left(  \mathfrak{a}_{1}-\mathfrak{A}_{1}\right)  }\ ,
\end{equation}
while, $r^{3}$ gives
\begin{equation}
\frac{\mathfrak{b}_{3}}{\mathfrak{b}_{1}}=2\frac{\mathfrak{A}_{2}%
^{2}\mathfrak{a}_{1}^{2}-\mathfrak{A}_{1}^{2}\mathfrak{a}_{2}^{2}%
+\mathfrak{A}_{1}\mathfrak{a}_{1}\left(  \mathfrak{A}_{1}\mathfrak{a}%
_{3}-\mathfrak{A}_{3}\mathfrak{a}_{1}\right)  }{\mathfrak{A}_{1}%
^{2}\mathfrak{a}_{1}^{2}\left(  \mathfrak{a}_{1}^{2}-\mathfrak{A}_{1}%
^{2}\right)  }\ ,
\end{equation}
and finally, $r^{4}$ gives the unwieldy expression%
\begin{equation}
\frac{\mathfrak{b}_{4}}{\mathfrak{b}_{1}}=\frac{\left(
\begin{array}
[c]{c}%
2\mathfrak{A}_{1}^{3}\mathfrak{A}_{2}\mathfrak{a}_{1}\mathfrak{a}_{2}%
^{2}+2\mathfrak{A}_{1}^{2}\mathfrak{A}_{3}\mathfrak{a}_{1}^{3}\mathfrak{a}%
_{2}-2\mathfrak{A}_{1}^{3}\mathfrak{A}_{2}\mathfrak{a}_{1}^{2}\mathfrak{a}%
_{3}-2\mathfrak{A}_{1}\mathfrak{A}_{2}^{2}\mathfrak{a}_{1}^{3}\mathfrak{a}%
_{2}-8\mathfrak{A}_{1}^{3}\mathfrak{a}_{1}^{2}\mathfrak{a}_{2}\mathfrak{a}%
_{3}+8\mathfrak{A}_{1}^{2}\mathfrak{A}_{2}\mathfrak{A}_{3}\mathfrak{a}_{1}%
^{3}\\
-7\mathfrak{A}_{1}^{4}\mathfrak{a}_{1}\mathfrak{a}_{2}\mathfrak{a}%
_{3}+7\mathfrak{A}_{1}\mathfrak{A}_{2}\mathfrak{A}_{3}\mathfrak{a}_{1}%
^{4}+3\mathfrak{A}_{1}^{4}\mathfrak{a}_{1}^{2}\mathfrak{a}_{4}+3\mathfrak{A}%
_{1}^{3}\mathfrak{a}_{1}^{3}\mathfrak{a}_{4}-3\mathfrak{A}_{1}^{3}%
\mathfrak{A}_{4}\mathfrak{a}_{1}^{3}-3\mathfrak{A}_{1}^{2}\mathfrak{A}%
_{4}\mathfrak{a}_{1}^{4}\\
+5\mathfrak{A}_{1}^{3}\mathfrak{a}_{1}\mathfrak{a}_{2}^{3}-5\mathfrak{A}%
_{1}\mathfrak{A}_{2}^{3}\mathfrak{a}_{1}^{3}+4\mathfrak{A}_{1}^{4}%
\mathfrak{a}_{2}^{3}-4\mathfrak{A}_{2}^{3}\mathfrak{a}_{1}^{4}+\mathfrak{A}%
_{1}^{3}\mathfrak{A}_{3}\mathfrak{a}_{1}^{2}\mathfrak{a}_{2}-\mathfrak{A}%
_{1}^{2}\mathfrak{A}_{2}\mathfrak{a}_{1}^{3}\mathfrak{a}_{3}-\mathfrak{A}%
_{1}^{2}\mathfrak{A}_{2}^{2}\mathfrak{a}_{1}^{2}\mathfrak{a}_{2}%
+\mathfrak{A}_{1}^{2}\mathfrak{A}_{2}\mathfrak{a}_{1}^{2}\mathfrak{a}_{2}^{2}%
\end{array}
\right)  }{\mathfrak{A}_{1}^{3}\mathfrak{a}_{1}^{3}\left(  \mathfrak{a}%
_{1}^{2}-\mathfrak{A}_{1}^{2}\right)  \left(  \mathfrak{a}_{1}^{2}%
+\mathfrak{a}_{1}\mathfrak{A}_{1}+\mathfrak{A}_{1}^{2}\right)  }\ .
\end{equation}
And so it goes. \ We obtain $\mathfrak{b}_{j}/\mathfrak{b}_{1}$ as functions
of the $\mathfrak{a}$s and $\mathfrak{A}$s. \ But again, it is overly
ambitious to carry the series approximation for $\beta$ beyond that for either
$u$ or $\sigma\left(  u\right)  $. \ 

There remains only one coefficient to be determined, namely, $\mathfrak{b}%
_{1}$. \ This is given by \
\begin{equation}
\mathfrak{b}_{1}=\ln\lambda=\ln\left(  \frac{\mathfrak{a}_{1}\left(  L\right)
}{\mathfrak{a}_{1}\left(  \ell\right)  }\right)  \ . \label{b1}%
\end{equation}
Although repetitious, a complete derivation of this result goes as follows.
\ Setting the scale of $t$ so that $u\left(  t=1\right)  =\sigma\left(
u\right)  $, Schr\"{o}der's equation is, once again,%
\begin{equation}
\Psi\left(  \sigma\left(  u\right)  \right)  =\lambda\Psi\left(  u\right)  \ ,
\label{SchroederYetAgain}%
\end{equation}
and from $u\left(  t\right)  =\Psi^{-1}\left(  \lambda^{t}\Psi\left(
u\right)  \right)  $ with $\frac{du\left(  t\right)  }{dt}=\beta\left(
u\left(  t\right)  \right)  $, we obtain $\beta\left(  u\left(  t\right)
\right)  =\left(  \ln\lambda\right)  \Psi\left(  u\left(  t\right)  \right)
/\Psi^{\prime}\left(  u\left(  t\right)  \right)  $. \ Now, at a fixed point
$u_{\ast}=\sigma\left(  u_{\ast}\right)  $, with $\Psi\left(  u_{\ast}\right)
=0$,\ a series solution of (\ref{SchroederYetAgain}) in powers of $\left(
u-u_{\ast}\right)  $ requires%
\begin{equation}
\lambda=\left.  \frac{d\sigma\left(  u\right)  }{du}\right\vert _{u=u_{\ast}%
}=\left.  \frac{d\sigma/dr}{du/dr}\right\vert _{r=0}=\frac{\mathfrak{a}%
_{1}\left(  L\right)  }{\mathfrak{a}_{1}\left(  \ell\right)  }\text{ .}%
\end{equation}
and also that $\beta\left(  u\right)  =\left(  \ln\lambda\right)  \left(
u-u_{\ast}\right)  +O\left(  \left(  u-u_{\ast}\right)  ^{2}\right)  $.
\ Hence (\ref{b1}) is obtained. \ Alternatively, in terms of the original
series involving the bare lattice coupling, we have%
\begin{equation}
\lambda=\frac{1+\sum_{j\geq2}\left(  j-1\right)  c_{j}\left(  L\right)
g_{0\ast}^{2j}}{1+\sum_{j\geq2}\left(  j-1\right)  c_{j}\left(  \ell\right)
g_{0\ast}^{2j}}=\frac{u_{\ast}+\sum_{j\geq1}j~c_{j}\left(  L\right)  g_{0\ast
}^{2j-1}}{u_{\ast}+\sum_{j\geq1}j~c_{j}\left(  \ell\right)  g_{0\ast}^{2j-1}%
}\ .
\end{equation}

\end{document}